\documentclass[twocolumn]{aastex631}
\usepackage{amsmath}
\newcommand{\hl}[1]{#1}

\shorttitle{DI-Her Misalignment}
\shortauthors{}
\graphicspath{{./}{figures/}}

\begin{document}

\title{DI Herculis Revisited: Starspots, Gravity Darkening, and 3-D Obliquities}

\author[0000-0002-1001-1235]{Yan Liang}
\affiliation{Department of Astrophysical Sciences, Princeton University,
Princeton, NJ 08544, USA}
\author[0000-0002-4265-047X]{Joshua N.\ Winn}
\affiliation{Department of Astrophysical Sciences, Princeton University,
Princeton, NJ 08544, USA}
\author[0000-0003-1762-8235]{Simon H.\ Albrecht}
\affil{Stellar Astrophysics Centre, Department of Physics and Astronomy, Aarhus University, Ny Munkegade 120, 8000 Aarhus C, Denmark}



\begin{abstract}

DI~Herculis is an eclipsing binary famous for a longstanding disagreement between
theory and observation of the apsidal precession rate, which was resolved when both stars
were found to be severely misaligned with the orbit.
We used data from the Transiting Exoplanet Survey Satellite (TESS) to
refine our knowledge of the stellar obliquities and
sharpen the comparison between the observed and theoretical precession rates.
The TESS data show variations with a 1.07-day period,
which we interpret as rotational modulation from starspots on the primary star.
This interpretation is supported by the detection of photometric
anomalies during primary eclipses consistent with starspot crossings.
The secondary eclipse light curve shows a repeatable asymmetry which
we interpret as an effect of gravity darkening.
By combining the TESS data with previously obtained data, we determined
the three-dimensional spin directions of both stars.
Using this information, the updated value of the theoretical
apsidal precession rate (including the effects of tides, rotation, and general
relativity) is $1.35^{+0.58}_{-0.50}$~arcsec/cycle.
The updated value of the observed rate (after including new TESS eclipse times)
is \hl{$1.41^{+0.39}_{-0.28}$}~arcsec/cycle.
Given the agreement between the observed and theoretical values,
we fitted all the relevant data simultaneously assuming the
theory is correct. This allowed us to place tighter constraints
on the stellar obliquities, which are
\hl{$75^{+3}_{-3}$ and $80^{+3}_{-3}$} degrees for the primary and secondary
stars, respectively.
\vspace{-0.1in}
\end{abstract}

\keywords{Detached binary stars (375), Eclipsing binary stars (444), Gravity darkening (680)}

\section{Introduction}
\label{sec:intro}

For several decades, the observed apsidal precession rate of the
eclipsing binary DI\,Herculis (B4V+B5V, $P=10.55$~days, $e=0.50$)
appeared to be too slow to be consistent with theoretical expectations
\citep{martynov1980relativistic,guinan1985apsidal,claret1998some}.
Many hypotheses were discussed, including the existence of a circumbinary planet,
a distant third star, a failure of general relativity, and a misalignment between the spin and orbital axes.
\cite{albrecht2009misaligned} proved the last of these hypotheses to be correct,
by observing the Rossiter-McLaughlin effect.
However, the true obliquities of the two stars --- which are needed to calculate the theoretical
precession rate --- were poorly constrained by the observations.
This is because the Rossiter-McLaughlin effect is mainly sensitive
to the sky-projected obliquity $\lambda$, and insensitive to the inclination $i$ of
the rotation axis with respect to the line of sight.
Because of the need to marginalize over the unknown inclinations of both stars,
the theoretical apsidal precession rate was subject to a large uncertainty
\citep{albrecht2009misaligned,claret2010di,Claret+2021}.

\begin{figure*}[ht!]
\epsscale{1.1}
\plotone{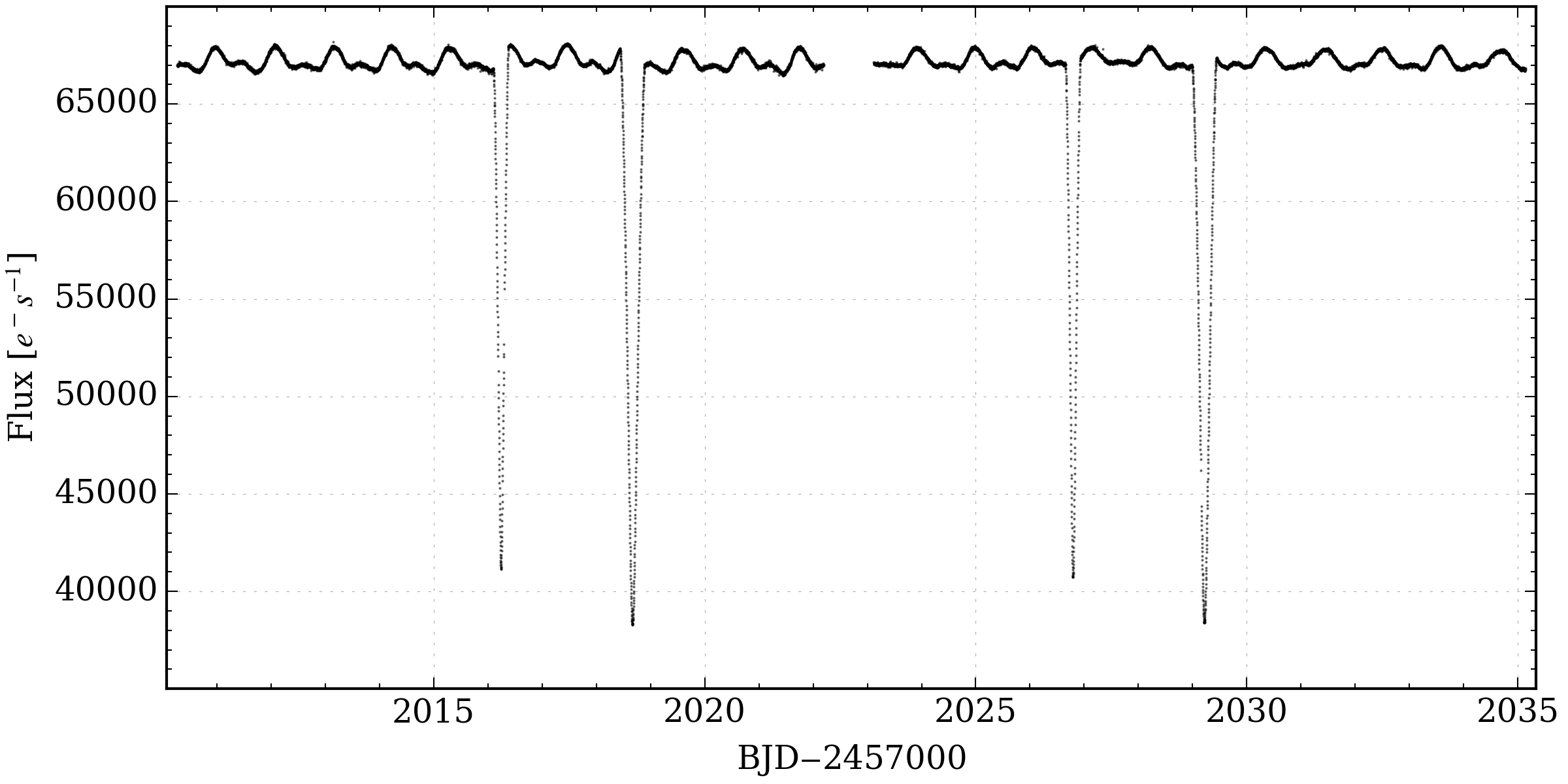}\caption{TESS observation of DI\,Her with 2-min time sampling,
covering two orbital periods. \label{fig:tess}}
\end{figure*}

One way to improve our knowledge of the stars' orientations is to detect and model
the effects of gravity darkening on the eclipse light curves.
Gravity darkening refers to the equator-to-pole variation in the intensity of a star's photosphere
due to rotation. The equatorial zone is centrifugally lifted to higher elevation, resulting in a lower
effective temperature and a lower emergent intensity.
\hl{For an idealized
radiative star, the local effective temperature varies with the local effective gravitational acceleration as $T_{\rm eff}\propto g_{\rm eff}^{0.25}$,
a result known as the \mbox{\cite{von1924radiative}} theorem.}
The resulting intensity variations with stellar latitude cause changes in the eclipse light curve
which depend on both $\lambda$ and $i$.
\cite{philippov2013analysis} developed this technique and attempted to
measure the three-dimensional obliquities of both stars of DI\,Herculis. However,
the only available light curves were ground-based and had incomplete
coverage of the eclipses, which limited their ability to 
improve on the system parameters.
Instead, they took the reverse approach: they determined the obliquities by requiring agreement
between the theoretical and observed apsidal precession rates.

In this paper, we investigated the time-series photometry of DI~Herculis
from the NASA Transiting Exoplanet Survey Satellite (TESS; \citealt{Ricker+2015}).
Our original goal was to detect the effects
of gravity darkening on the eclipse light curves. After inspecting the data,
we also saw the potential to measure the rotation period of one or both stars, given
the obvious pattern of quasiperiodic flux modulation (Section~\ref{sec:rot}).
The secondary eclipse light curves showed the expected asymmetry due to gravity
darkening, allowing us to determine the secondary star's obliquity
(Section~\ref{sec:sec-ecl}).
The primary eclipse light curves showed larger and non-repeating
anomalies, probably due to the same starspots that produce the quasiperiodic
flux modulation, which prevented a clear detection of gravity
darkening (Section~\ref{sec:prim-ecl}).
Nevertheless, by combining information from the gravity darkening of the secondary star and
the rotation period of the primary star, along with previous measurements
of the Rossiter-McLaughlin effect and the stars' projected rotation velocities,
we were able to determine the three-dimensional obliquities of both
stars with improved precision (Section~\ref{sec:time-evol}).
We also used the new TESS eclipse timings to update the measurement of the
observed apsidal precession rate (Section~\ref{sec:aps-obs}).
This allowed us to perform a more stringent comparison between the
theoretical and observed apsidal precession rates (Section~\ref{sec:aps-calc}).
Finally, we fitted all the available data to refine our
estimates of the key system parameters (Section~\ref{sec:synthesis}).

\section{Rotational modulation}
\label{sec:rot}

DI Herculis was observed by TESS between 2020~June~9 and 2020~July~4, within Sector 26 of the Prime Mission.
We downloaded the light curve with 2-minute time sampling that had been prepared
by the TESS Science Processing Operations Center (SPOC). Shown in Figure~\ref{fig:tess},
the light curve encompasses two primary eclipses and two secondary eclipses.  There is also quasiperiodic
variability with an amplitude of $\sim$1\% and a period of approximately one day.
The variations resemble those of rotating starspots on the stellar photosphere, and unlike those of pulsation.
While slowly pulsating B stars sometimes have periods as long as one day, the light
curves usually show strong beating between multiple frequencies \citep[see, e.g.,][]{aerts2006discovery}. In contrast,
DI\,Herculis shows two maxima per cycle, with amplitudes that vary slowly
compared to the period. This is typical of stars with several persistent spots.

\begin{figure}[ht!]
\epsscale{1.15}
\plotone{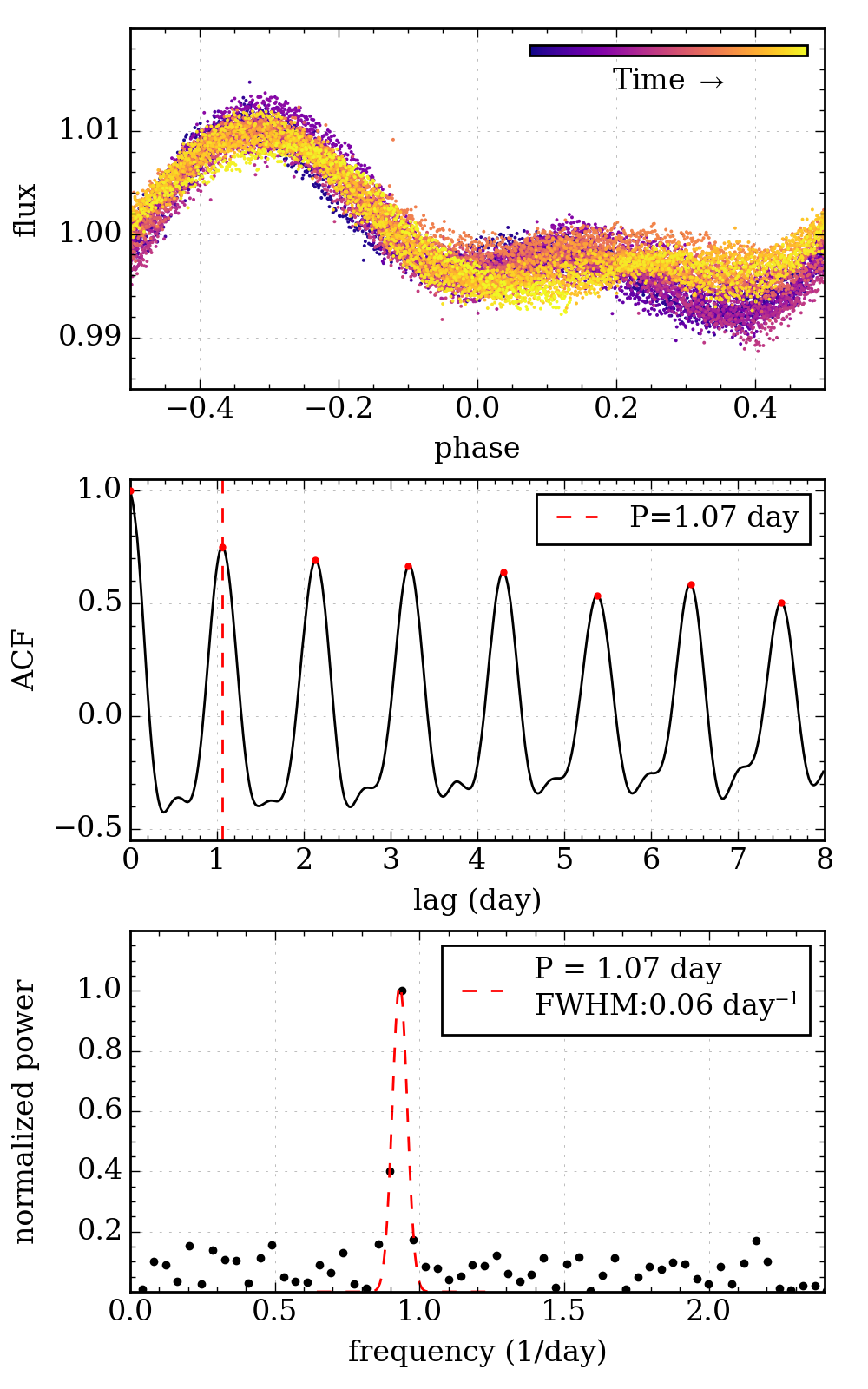}
\caption{{\it Top}: Phase-folded and normalized TESS light curve,
color-coded by time.
The features in the first half of the light curve were more stable than those in
the second half.
{\it Middle}: Autocorrelation function.
{\it Bottom}: Fourier transform.
\label{fig:acf}}
\end{figure}

To measure the period, we calculated the autocorrelation function (ACF) of the light curve,
following the approach of \cite{2013MNRAS.432.1203M}.
Given a time-domain signal $\{x_i\}(i=1,2,...,N)$ with uniform
spacing $\Delta t$,
the auto-correlation coefficient $r$ at lag $\tau_k = k\Delta t$ is defined as
\begin{equation}
r(\tau_k) = \frac{\sum_{i=1}^{N-k} (x_i-\bar{x})(x_{i+k}-\bar{x})}{\sum_{i=1}^{N} (x_i-\bar{x})^2}.
\end{equation}
Before computing the ACF, we replaced the data values obtained during eclipses with the mean out-of-eclipse flux.
The ACF is shown in the middle panel of Figure~\ref{fig:acf}.
We fitted the peak positions with a linear function of peak number, obtaining a slope
of $1.07\pm 0.01$~days.
The Fourier transform of the light curve, shown in the bottom panel, also shows a peak
corresponding to a period of $1.07$~days. The top panel shows the light curve folded
with a 1.07-day period.
The identification of the photometric period with the stellar rotation period is subject to additional uncertainty
beyond the formal uncertainty of 0.01~day, because of the limited
coherence of the signal and systematic effects due to differential rotation.
To be conservative, for our subsequent analysis we adopted $P_{\rm rot} = 1.07\pm 0.10$~days.

Which star is rotating with this period?
It is possible that both stars
contribute to the observed rotational modulation;
given the stars' similar masses, we expect them 
to have similar rotation periods. 
However, later in this paper we will argue that the primary star
is the dominant contributor to the observed flux modulation, based on the pattern of
anomalies that were seen in the primary eclipse light curves (\S~\ref{sec:prim-ecl}).


\begin{figure}[ht!]
\epsscale{1.15}
\plotone{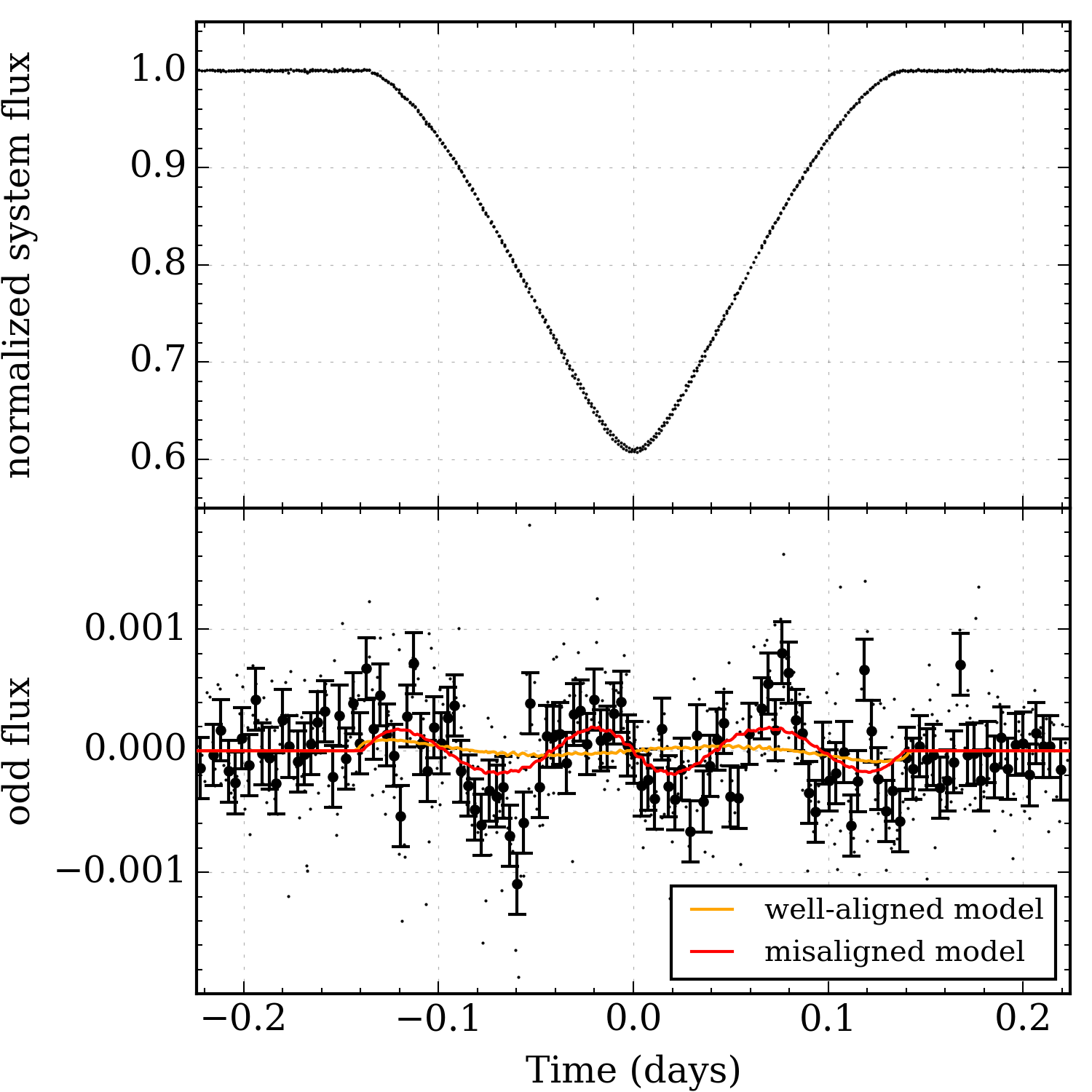}
\caption{{\it Top}: Phase-folded secondary eclipse light curve (black points).
{\it Bottom}: The irreducible antisymmetric component.
The black points with error bars are time-averaged data. The orange and
red curves represent the well-aligned and misaligned models, respectively.
\label{fig:irr}}
\end{figure}

\begin{figure*}[ht!]
\plotone{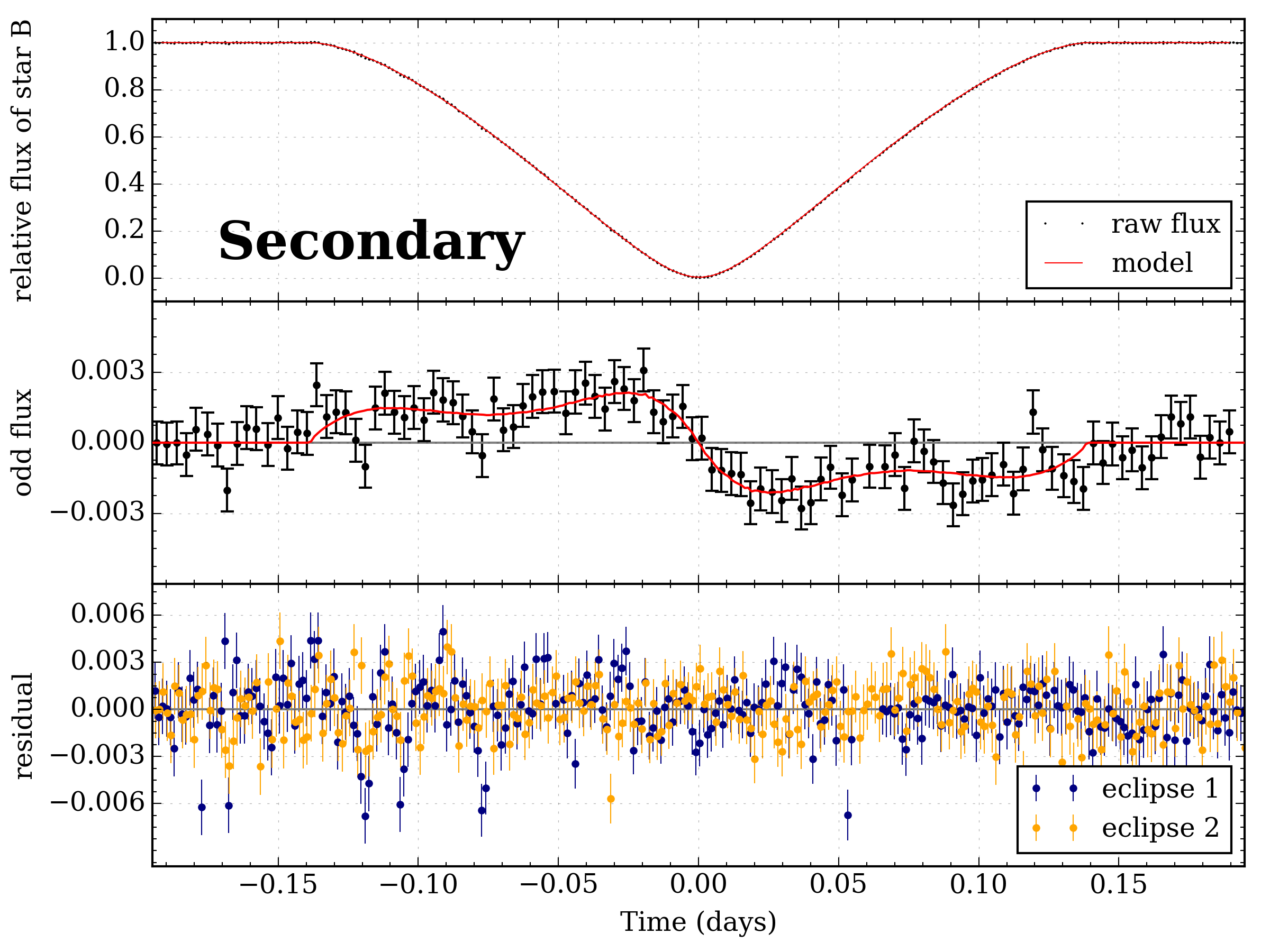}
\caption{TESS data for secondary eclipses.
{\it Top}: Phase-folded light curve (black points) and the best-fitting model (red curve),
after subtracting the modeled contribution of the primary star and setting the
flux to unity outside eclipses.
{\it Middle}: The irreducible antisymmetric component. The darker points with error bars are time-averaged data. We attribute the asymmetry to the effects of gravity darkening.
{\it Bottom}: Residuals, after subtracting the best-fitting model (including
gravity darkening) from the data.
\label{fig:lc2}}
\end{figure*}

\section{Secondary eclipses}
\label{sec:sec-ecl}

We begin by presenting the secondary eclipse light curves, because they
turned out to be more straightforward to interpret than the primary
eclipses.
The top panel of Figure~\ref{fig:irr} shows the phase-folded secondary eclipse light curve
based on the two TESS eclipses, after normalizing the data
to have unit flux outside of the eclipses.  Normalization was achieved
by dividing the data by a cubic function of time that
had been fitted to the out-of-eclipse data within 1.4~hr of each eclipse.

\hl{Ordinarily, we expect eclipse light curves to be symmetric about the time of mid-eclipse,
apart from effects of orbital eccentricity, Doppler shifts, and relativistic beaming,
which are usually too small to observe.}
Gravity darkening introduces an asymmetry between the first and second halves
of the eclipse, except for the special cases $\lambda=0^\circ$, $\pm 90^\circ$,
and $180^\circ$ \citep[see, e.g.,][]{Barnes2009,masuda2015spin}.
For this reason, prior to any parametric modeling, we extracted the
antisymmetric component of the light curve as a simple diagnostic
of gravity darkening. 

The antisymmetric component of a normalized light curve $f(\tau)$ is
defined as
\begin{equation}
\label{eq:odd}
    f_{\text{odd}}(\tau) = \frac{f(\tau)-f(-\tau)}{2},
\end{equation}
where $\tau=t-t_0$, the time relative to the mid-eclipse time.
Given a value of $t_0$, we used linear
interpolation to estimate $f(-\tau)$ for each data point
$f(\tau)$.  Because $t_0$ is not known {\it a priori}, we
iterated to find the value of $t_0$ that minimizes
the sum of squares of $f_{\text{odd}}(\tau)$.
We refer to the antisymmetry that remains even after allowing
for freedom in $t_0$ as the ``irreducible'' antisymmetric
component, which is shown in the bottom panel of Figure~\ref{fig:irr}. 

Both eclipses showed the same wavy pattern of residuals. The flux
at mid-ingress was 0.1\% lower than at mid-egress.
We interpret this asymmetry
as the expected effect of gravity darkening. (Another source of light-curve
asymmetry is orbital eccentricity, but we confirmed
through numerical light-curve modeling that the effects of eccentricity
are too small to explain the observed antisymmetry.) To model the light curve,
we used the \texttt{PyTransit} code by \cite{Parviainen2015}.
This code was developed for modeling planetary transits.
We made a small adjustment to the code to account for the light
from the eclipsing star and any other sources within the photometric aperture.\footnote{\hl{This
code does not account for tidal and reflection effects.
We used the more sophisticated PHOEBE code \mbox{\citep{conroy2020physics}} to confirm that the expected
size of these effects is below the noise level of the TESS light curves.
The possible exception is reflected light near secondary eclipse,
which may be comparable to the noise level. Nevertheless, we decided
to neglect this effect because the reflected-light signal is nearly symmetric around the mid-eclipse and occurs over a longer timescale than the eclipses,
and therefore does not have a major impact on our eclipse model.}}

The total observed flux can be written
\begin{align}
\label{eq:total}
    F(t) &= \sum_{j=1}^2 \left[ F_{j,\text{full}}
    - F_{j,\text{ecl}}(t)\right] + F_3(t) \\
         &= F_{\text{out}}(t)-\sum_{j=1}^2 F_{j,\text{ecl}}(t).
\end{align}
In the first line, $F_{j,\text{full}}$ is the ``full flux'' of star $j$,
$F_{j,\text{ecl}}(t)$ is the flux blocked by the eclipsing star,
and $F_3(t)$ represents any ``third light'' from another star or imperfectly subtracted background light in the images.
In the second line, all of the contributions except the eclipse-specific variation
have been lumped together to define $F_\text{out}(t)$.
After doing so, the normalized eclipse light curve can be written
\begin{equation}
    f_{j,\text{ecl}}(t) = \frac{F_{j,\text{full}}-F_{j,\text{ecl}}(t)}{F_{j,\text{full}}} =1+\frac{F(t)-F_\text{out}(t)}{F_{j,\text{full}}}.
\end{equation}
Since the information we seek is within the function
$F_{j,\text{ecl}}(t)$, we took $F_{\rm out}(t)$
to be an {\it ad hoc} cubic function of time.
Thus, there were 5 adjustable parameters in the normalization
of each eclipse:
$F_{j,\text{full}}$ and the four coefficients of the cubic function.
\hl{Note that $F_{j,\text{full}}$ and $F_\text{out}(t)$
implicitly provide the same information as the ratio of stellar temperatures.}

\begin{figure*}[ht!]
\plotone{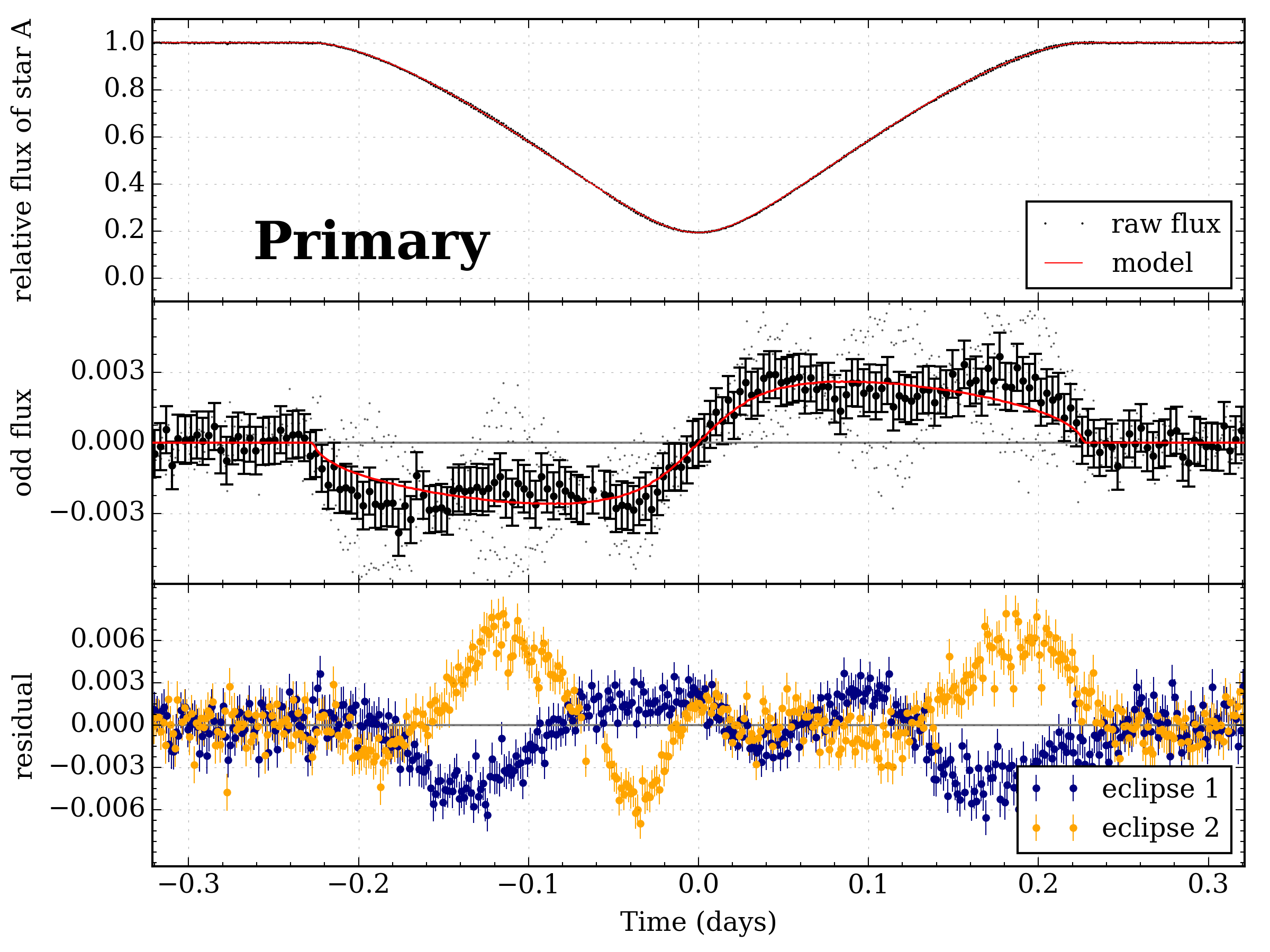}
\caption{TESS data for primary eclipses.
{\it Top}: Phase-folded light curve (black points) and the best-fitting model (red curve), normalized by the full primary flux.
{\it Middle}: Antisymmetric component. The darker points with error bars are time-averaged data.
{\it Bottom}: Residuals, after subtracting the best-fitting model from the data. The residuals
differ for the two eclipses, leading us to attribute the asymmetries and other anomalies
to starspot crossings rather than gravity darkening.
\label{fig:lc1}}
\end{figure*}

The parameters of the model can be divided into three groups.
First are the usual eclipse-specific parameters:
the stellar radius ratio ($R_1/R_2$),
the ratio of the semimajor axis to the secondary star's radius ($a/R_2$),
the orbital inclination ($i_o$),
the two coefficients of a quadratic limb-darkening law ($u_1$ and $u_2$),
and a particular mid-eclipse time ($T_{0,2}$).
Second are the orbital parameters:
the period ($P_\text{orb}$),
eccentricity ($e$),
and argument of periapsis ($\omega$).
Third are the parameters specific to gravity darkening:
the rotation period of the secondary star ($P_\text{rot,2}$),
the stellar inclination ($i_{s,2}$),
the projected obliquity ($\lambda_2$),
the average density ($\rho_2$),
the pole temperature ($T_\text{pole,2}$),
and the gravity-darkening coefficient ($\beta_{g,2}$).

Due to the spin and orbital precession, the stellar inclination ($i_{s,2}$), the projected obliquity ($\lambda_2$), and the argument of periapsis ($\omega$) vary on a timescale of years. \hl{ In the light curve model, these parameters are assumed to be constant in time over 25 days of the TESS observations.}

The local temperature at latitude $\phi$ is specified by
\begin{equation}
    T(\phi)=T_\text{pole}\left[\frac{g(\phi)}{g_\text{pole}}\right]^{\beta_g}
\end{equation}
where $T_\text{pole}$ is the pole temperature, $g_\text{pole}$ is the pole gravity, $g(\phi)$ is the local gravity, and $\beta_{g}$ is the gravity-darkening coefficient. \texttt{PyTransit} calculates the passband-integrated emergent flux from each point on the star as a function of temperature.
Since the secondary eclipse light curve is insensitive to the pole temperature, we assumed $T_{\text{pole},2} \approx T_{\text{eff},2}$ and adopted the value 15{,}400\,K from \citep{claret2010di}.
We imposed Gaussian priors on the eccentricity ($e=0.489\pm0.030$) and 
argument of peripasis ($\omega=329.9\pm0.6^\circ$) based on the radial-velocity
analysis of \cite{popper1982rediscussion}.
We adopted limb-darkening coefficients $u_1=0.124\pm0.1,u_2=0.262\pm0.1$ and the gravity-darkening coefficient $\beta_{g,2}=0.086\pm0.025$ from the work by \cite{claret2017limb}. 
Uniform priors were employed for all other parameters.

The bottom panel of Figure~\ref{fig:lc2} shows the residuals between the
data and the best-fitting model.
Using the formal flux uncertainties from the SPOC light curve,
we found the minimum $\chi^2$ to be 1246 with 538 degrees of freedom.
This is probably due to a combination
of underestimated flux uncertainties, and departures of the true
limb-darkening and gravity-darkening profiles from our idealized
models. To provide more realistic parameter uncertainties,
we enlarged the flux uncertainties by a factor of 1.5,
which reduces $\chi^2$ to equal the number of degrees of freedom.


We used a Markov Chain Monte Carlo method to
determine the posterior distribution for all the parameters. 
For the parameters of greatest interest, the marginalized posterior
distributions led to results of
$P_\text{rot,2}=0.89^{+0.23}_{-0.16}$~days,
$i_{s,2}=106^{+19}_{-14}$~degrees,
and $\lambda_{2}=38^{+26}_{-14}$ degrees. This result for $\lambda_{2}$ is about
1.7-$\sigma$ away from the previous result of $\lambda_2 = 84\pm 8$~degrees obtained
by fitting the Rossiter-McLaughlin effect \citep{albrecht2009misaligned}.
The uncertainties in the parameters are strongly correlated.
In particular, the credible interval for $P_\text{rot,2}$ posterior extends to values as long
as $\sim$3~days. 
Thus, although the light curve is well-fitted by the model,
there are large uncertainties and strong degeneracies.
Section~\ref{sec:time-evol} describes our effort
to incorporate more external information
to better constrain the star's orientation and rotation period.

\section{Primary eclipses}
\label{sec:prim-ecl}

Figure~\ref{fig:lc1} shows the phase-folded primary eclipse light curve, the irreducible
antisymmetric component of each of the two primary eclipses, and the residuals between the data and the best-fitting model.
In this case, the two eclipses showed different antisymmetric components,
and the residuals for the first eclipse showed a different 
pattern than those for the second eclipse.
Thus, gravity darkening cannot be solely responsible for the
light-curve asymmetries.

Instead, we think the anomalies in the primary eclipses were produced when the
eclipsing star crossed over dark spots on the primary star.  This seems reasonable
given that the TESS light curve showed quasiperiodic modulation indicative of
starspots. As a proof of concept, we 
modified \texttt{PyTransit} to include spots.
We assumed each spot to be circular
and have a uniform effective temperature, for simplicity.
A spot is specified by 5 parameters:
the radius $r/R_\star$,
the latitude $\phi_0$ and longitude $l_0$ at a reference epoch $t_{\rm ref}$,
and the factor $\epsilon$ by which the effective temperature is reduced relative to the surrounding photosphere. We chose $t_{\rm ref} = t_0$, the time of mid-eclipse.

The description of the light curve needed to be modified to account
for the loss of light due to spots. Instead of Equation~\ref{eq:total},
we used
\begin{align}
    F(t) &= \sum_{j=1}^2 \left[ F_{j,\text{full}}-F_{j,\text{spot}}(t)-F_{j,\text{ecl}}(t)\right] + F_3(t) \\
         &= F_{\text{out}}(t)-\sum_{j=1}^2 F_{j,\text{ecl}}(t).
\end{align}
Here, $F_{j,\text{spot}}(t)$ is the loss of flux due to spots.
The loss of light during eclipses, $F_{j,\text{ecl}}(t)$,
also depends on the starspot pattern. When the blocked portion
of the eclipsed star contains a dark spot, the loss of light is
not as large as it would have been without the spot.

\begin{figure}[ht!]
\epsscale{1.15}
\plotone{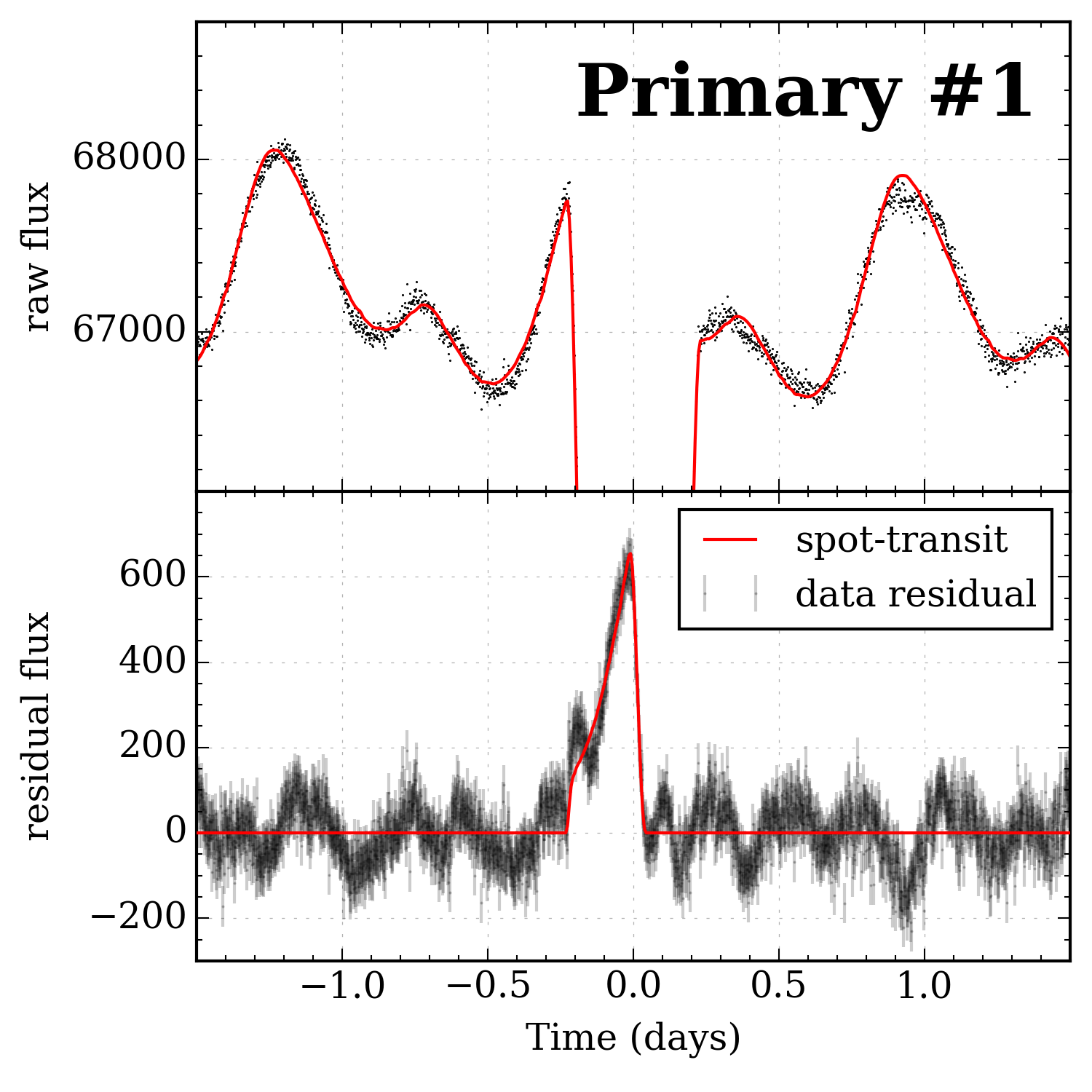}
\caption{Spot modeling of the primary eclipse.
{\it Top}: TESS data (black dots) and the best-fitting model including two
spots (red curve).
{\it Bottom}: Residuals between the data and the same best-fitting model
without spot-transits. The red curve shows the spot-model prediction.
\label{fig:prim}}
\end{figure}

\begin{deluxetable}{lcc}[ht!]
\tablenum{1}
\tablecaption{Spot parameters\label{tab:spot}}
\tablewidth{0pt}
\tablehead{\colhead{Parameter} &  \colhead{spot 1} & \colhead{spot 2}}
\startdata
$\epsilon$ & 0.75 & 0.77 \\
$r/R_\star$ & 0.37 & 0.31 \\
$l_0$ [deg] & 165 & 301 \\
$\phi$ [deg] & 68 & 11 \\
\enddata
\tablecomments{Parameters used to generate Figures \ref{fig:prim} and \ref{fig:LT}.}
\end{deluxetable}

We were able to achieve a satisfactory fit to the first primary eclipse light curve
using two spots. The model also includes the effects
of gravity darkening.
Figure~\ref{fig:prim} shows the TESS data surrounding the first primary
eclipse. The top panel focuses on the out-of-eclipse quasiperiodic modulation, and the
bottom panel shows the contribution due to spot crossings. The model provides
a reasonable fit, in both cases.
Thus, in the model, the same two spots that are responsible for
the out-of-eclipse modulation are also responsible for the eclipse anomalies.
This is why we think the primary star is the dominant contributor to the
quasiperiodic flux modulation, as alluded to earlier.
Table~\ref{tab:spot} gives the spot parameters, and the top
two panels of Figure~\ref{fig:LT} show the
model intensity and temperature
distribution across the stellar disk. 

We performed a similar exercise for the second primary eclipse, although
it did not seem worthwhile to pursue in detail.
Given the oversimplifications of the spot model,
the spot parameters are of limited interest.
Because of the uncertainties in spot modeling,
we decided not to use the primary eclipse light curve to constrain
the spin orientation of the primary star.


\begin{deluxetable}{lccc}[ht]
\tablenum{2}
\tablecaption{$v\sin{i}, \lambda$ measurements\label{tab:vsini}}
\tablewidth{0pt}
\tablehead{\colhead{Year} &  \colhead{$v_1\sin{i_{s,1}}$ [km/s]} & \colhead{$v_2\sin{i_{s,2}}$ [km/s]}& \colhead{Ref}}
\startdata
1974 & $34 \pm 6$ & $50 \pm 7$ &  1\\
1985 & $50 \pm 30$ & $50 \pm 30$ & 1 \\
1988 & $61 \pm 4$ & $75 \pm 4$ & 1 \\
2008 & $108 \pm 4$ & $116 \pm 4$ & 2 \\
2016 & $116 \pm 3$ & $121 \pm 4$ & This study \\
 \hline
\\[0em]
 \hline\hline
\\[-1em]
Year & $\lambda_1$ [deg] & $\lambda_2$ [deg]& Ref\\
\\[-1em]
\hline
2008 & $-72 \pm 4$ & $84 \pm 8$ & 2 \\
\enddata
\tablecomments{
Ref 1: \cite{1989AJ.....97..216R}\\
Ref 2: \cite{albrecht2009misaligned}. Note that they referred to $\beta$, which is simply $-\lambda$.
}
\end{deluxetable}

\section{Time evolution of the spin and orbital axes}
\label{sec:time-evol}

\cite{albrecht2009misaligned} measured $\lambda_1$ and $\lambda_2$ based on observations
of the Rossiter-McLaughlin effect.  They also reported measurements of both stars'
projected rotation velocities obtained sporadically over the past few decades.
The velocities are observed to change over time because the stars' rotation
axes are precessing around the total angular momentum vector of the system.
We used this information to supplement the fit to the TESS secondary eclipse light curve,
which by itself led to large parameter uncertainties
and strong degeneracies.

To obtain another measurement of the projected rotation velocity,
we analyzed a spectrum of DI Herculis from 2016 May 30
acquired with the HARPS-N\footnote{High Accuracy Radial velocity Planet Searcher for the Northern hemisphere.} spectrograph
on the 3.6m Telescopio Nazionale Galileo at La Palma.
The spectrum has a resolution of 115{,}000 and covers the wavelength range from 383 to 690~nm.
The standard data reduction software was used to process the spectrum.
We determined the $v\sin{i}$ value for both components by fitting the
region surrounding the Mg\,II line at 4481\AA{} with a rotational
broadening kernel, assuming a linear limb-darkening law with $u=0.4$ for both stars. 
Table \ref{tab:vsini} gives the results, along with the other
available measurements.


We modeled the three dimensional spin precession of the system following the approach of
\cite{philippov2013analysis} (see also \citealt{barker1975gravitational,1989AJ.....97..216R}).
The spin precession equations are:
%
\begin{align}\label{eq:sl}
\begin{split}
    \frac{d{\bold{\hat{L}}}}{dt} &= \left[\sum_{i=p,s}\Omega_{Q,j} \cos{\psi_j}\bold{\hat{S}}_j \right]\times \bold{\hat{L}},     \\
\frac{ d\bold{\hat{S}}_j}{dt} &= \left[ \Omega_{G,j}-\Omega_{P,j} 3\cos{\psi_j} \right] \bold{\hat{L}}\times \bold{\hat{S}}_j,
\end{split}
\end{align}
where $\bold{L}$ is the orbital angular momentum,
$\bold{S}_j$ is the spin angular momentum of star $j$, and
$\psi_j = \cos^{-1} \left( \bold{\hat{S}}_j \cdot \bold{\hat{L}} \right)$ 
is the obliquity of star $j$.
The obliquity can be related to the projected obliquity
$\lambda_j$ and inclination $i_o$ via
\begin{equation}
\cos \psi_j = \cos i_{s,j} \cos i_o +\sin i_{s,j} \sin i_o \cos \lambda_j.
\end{equation}

The frequencies $\Omega_{Q,j}$ and $\Omega_{P,j}$ are the orbital and spin precession rates, respectively,
due to the rotationally-induced
stellar quadrupole. They can be calculated using Equations B12 and B14 of \citep{philippov2013analysis},
\begin{align}\label{eq:omega-QP}
\begin{split}
\Omega_{Q,j} &= -k_{2j}\frac{M_p+M_s}{M_j}\frac{\omega_j^2}{\Omega_K(1-e^2)^2}\left(\frac{R_j}{a}\right)^5, \\
\Omega_{P,j} &= \frac{k_{2j}}{3\eta_j}\frac{M_{j'}}{M_j}\frac{\omega_j}{(1-e^2)^{3/2}}\left(\frac{R_j}{a}\right)^3.
\end{split}
\end{align}
They depend on the orbital angular frequency ($\Omega_K$), the rotational angular frequencies
($\omega_{j}$), the moment-of-inertia constants ($\eta_j$), and the apsidal-motion constants
($k_{2j}$). We adopted the values of $\eta_j$ and $k_{2j}$ reported by \cite{Claret+2021} (see Table~\ref{tab:param}).
The geodesic spin precession rate $\Omega_{G,j}$ is much smaller than $\Omega_{P,j}$
and was neglected in our calculations.


Given a set of initial conditions, we integrated Equation \ref{eq:sl} to obtain the time evolution of $\bold{\hat{L}}$ and $\bold{\hat{S}}_j$, and calculated the expected values of $v \sin i$ and $\lambda$ throughout history.
This spin precession model only invokes rotational oblateness (not general relativity or tidal distortion) 
and does not take the observed apsidal precession rate as an input. The independence of this model from
the apsidal precession rate is important because it allowed us to use the results
to test for agreement between the theoretical and observed apsidal precession rates, as described in the next section.

The fitting statistic was the sum of $\chi^2_{\rm lc}$, the goodness-of-fit to the TESS light curve,
and
\begin{equation}\label{eq:chi_spec}
\begin{split}
    \chi^2_\text{prec}&=\sum_{k}^{10}\frac{\left[v\sin{i}_\text{obs}(t_k)-v\sin{i}_\text{calc}(t_k)\right]^2}{\sigma^2_{\text{obs},k}}\\
&+\sum_{i}^{2}\frac{\left[\lambda_\text{obs}(t_j)-\lambda_\text{calc}(t_j)\right]^2}{\sigma^2_{\text{obs},j}},
\end{split}
\end{equation}
where $k$ indexes the ten $v\sin i$ measurements, and $j$ indexes the two $\lambda$ measurements.

\begin{figure}[ht]
\epsscale{1.15}
\plotone{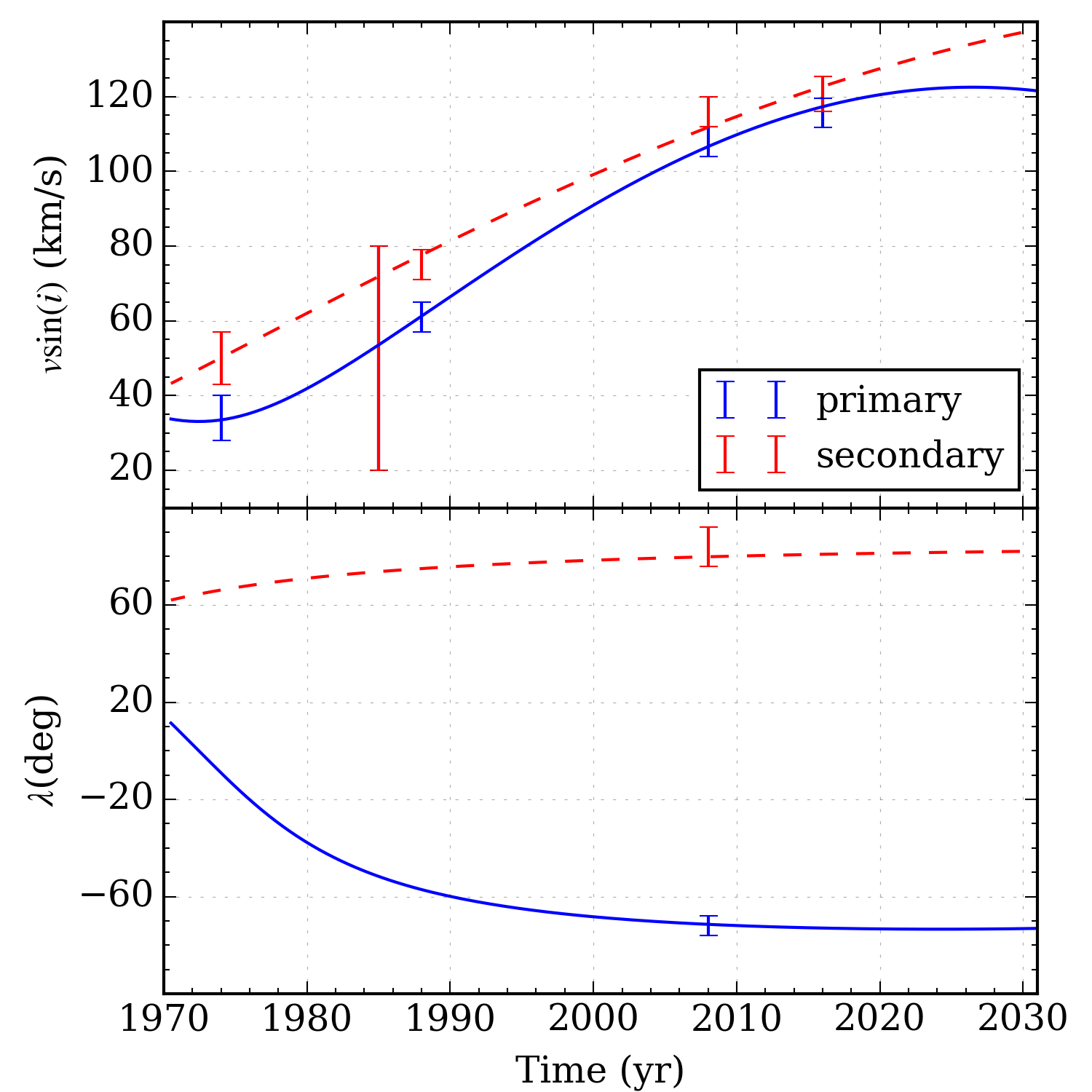}
\caption{Precession model.
{\it Top}: Measurements (data points) and best-fitting model (solid curves) of the time evolution
of $v\sin i$ for the primary (blue) and secondary (red).
{\it Bottom}: Same, for the projected obliquity $\lambda$.
\label{fig:vsini}}
\end{figure}

Instead of the lightcurve parameters $a/R_2$, $R_1/R_2$, and $\rho_2$,
for this more comprehensive model we used the parameters
$M_2$, $R_1$, and $R_2$. We also
introduced new parameters associated with the primary star: the mass $M_1$, inclination ($i_{s,1}$), rotation period ($P_{{\rm rot},1}$), and projected obliquity ($\lambda_1$).
We adopted Gaussian priors for the masses,
$M_1=5.1\pm0.2\,M_\odot$ and $M_2=4.4 \pm 0.2\,M_\odot$,
based on previous work by \cite{albrecht2009misaligned}.
We also placed a Gaussian prior of $1.07\pm 0.10$~days based
on the analysis presented in Section~\ref{sec:rot}.
The priors on the limb-darkening coefficients and gravity darkening coefficient were
the same as described in Section~\ref{sec:sec-ecl}.
We decided not to impose priors on $e$ and $\omega$, because the
TESS photometry should lead to tighter constraints on these parameters.

Figure~\ref{fig:vsini} shows the precession-related data and the best-fitting model, 
which has \hl{$\chi^2_\text{spec}=2.9$} and \hl{$\chi^2_\text{lc}=556$}, based
on 550 flux data points and 12 measurements of $v\sin i$ and $\lambda$, and 12 free parameters.  \hl{In total, $\chi^2_\text{tot}=559$} with 550 degrees of freedom. In the best-fitting model,
the spin precession periods for the primary and secondary stars are 215 and 418 years, respectively.


\section{Updated Measurement of Apsidal Precession Rate}
\label{sec:aps-obs}

The precise eclipse times derived from the TESS light curves give the opportunity to update
the measurement of the apsidal precession rate.
\hl{However, we decided to exclude the second primary eclipse from consideration, given that
our spot model provided a better fit to the first primary eclipse.
We combined the first TESS primary eclipse time and} the two TESS secondary eclipse times with the 61 eclipse timings reported by \cite{kozyreva2009pulsations} and \cite{claret2010di}.
In the few cases for which uncertainties were not reported,
we adopted an uncertainty of 0.001 day.

We measured the apsidal precession rate in the standard manner \citep{gimenez1983new}.
The model for the eclipse timings has six parameters:
the reference time ($T_0$),
the sidereal period ($P_{\rm s}$),
the orbital inclination ($i_o$),
the eccentricity ($e$),
the argument of periapsis at the reference epoch ($\omega_0$),
and the apsidal precession rate ($\dot \omega$). To improve the constraints, we used the photometric constraints on eccentricity and argument of periapsis obtained in section \ref{sec:time-evol} as Gaussian priors on $e$ and $\omega_0$. Uniform priors are assumed for the other parameters.

\begin{figure}[ht!]
\epsscale{1.15}
\plotone{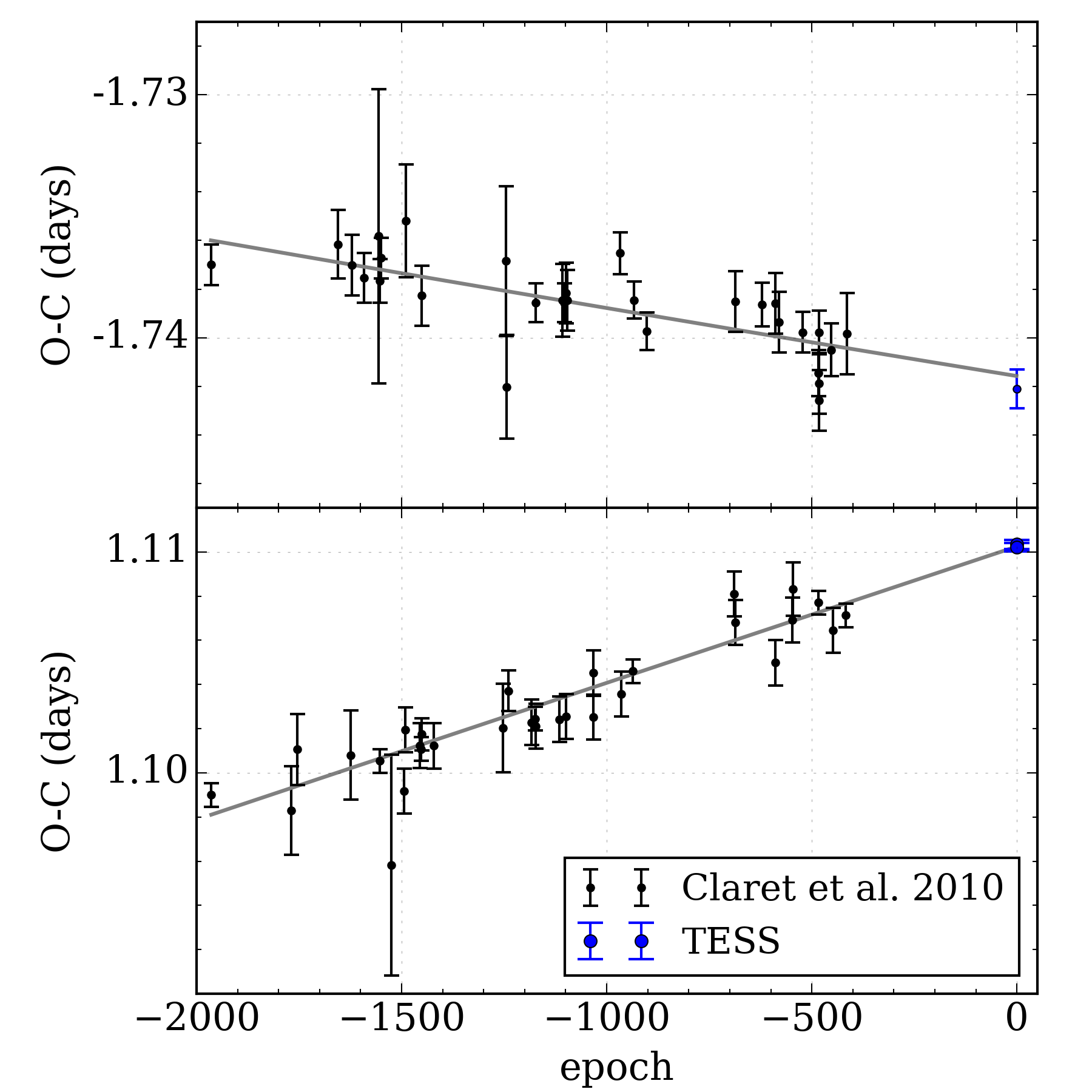}
\caption{
{\it Top}: Primary eclipse timing residuals, after subtracting $T_0+ EP_{\rm s}$, where $T_0$ is the reference time,
$E$ is the epoch number, and $P_{\rm s}$ is the sidereal period.
{\it Bottom}: Secondary eclipse timing residuals, after subtracting $T_0 + P_{\rm s}/2 +EP_{\rm s}$.
\label{fig:O-C}}
\end{figure}

The contribution to $\chi^2$ from the timings is
\begin{equation}\label{eq:chi_time}
    \chi^2_\text{timing}=\sum_{i=1}^{63}\frac{(T_{\text{obs},i}-T_{\text{calc},i})^2}{\sigma^2_{\text{obs},i}}.
\end{equation}
The best-fitting model is shown in Figure~\ref{fig:O-C}. \hl{Using the formal
uncertainties reported in the literature, we found $\chi^2 =132$ with 58 degrees of freedom. The primary eclipse timings contribute disproportionately
to $\chi^2_{\rm timing}$, probably because the previously obtained data were also affected by starspot anomalies. Specifically, $\chi^2_{\rm timing,p} =90.2$ for 28 degrees of freedom and $\chi^2_{\rm timing,s} =41.8$ for 30 degrees of freedom. 
To account for systematic errors in the timing measurements,
for fitting purposes we added a systematic uncertainty term in quadrature
with the formal uncertainties in order to force $\chi^2_{\rm timing} = N_{\rm dof}$.
The systematic error term was 0.0007~days for the primary
eclipses and 0.0002~days for the secondary eclipses.}
The apsidal precession rate was found to be
\hl{$\dot{\omega}_\text{obs}=1.41^{+0.39}_{-0.28}$}~arcsec/cycle.
This is consistent with the previously-determined value of
$\dot{\omega}_\text{obs}=1.51\pm0.43$ arcsec/cycle \citep{claret2010di}.

\section{Comparison between Observed and Theoretical Apsidal Precession Rates}
\label{sec:aps-calc}

The theoretical rate of apsidal precession arises from three different physical effects:
general relativity, tidal distortion, and rotational oblateness \citep{barker1975gravitational,Shakura1985,philippov2013analysis,claret2010di}:
\begin{equation}\label{eq:w1}
\dot \omega = \dot \omega_\text{GR} + \sum_{j=1}^2 \dot \omega_{\text{tidal},j}+ \dot \omega_{\text{rot},j}\phi_j,
\end{equation}
where $\phi_j$ depends on the the spin orientation:
\begin{align}
\begin{split}
\phi_j = &-\frac{1}{\sin^2 i_o}\cos\psi_j(\cos\psi_j-\cos i_{s,j}\cos i_o)\\
               &-\frac{1}{2}(1-5\cos^2\psi_j).
\end{split}
\end{align}
The individual contributions are:
\begin{align}
&\dot \omega_\text{GR}  = \frac{3 G (M_1+M_2)}{c^2a(1-e^2)} \\ 
&\dot \omega_{\text{tidal},j} = 15k_{2j}\frac{M_{j'}}{M_j} \frac{8+12e^2+e^4}{8(1-e^2)^5}\left[\frac{R_j}{a}\right]^5 \\
&\dot \omega_{\text{rot},j} = \frac{P^2_\text{orb}}{P^2_{\text{rot},j}}
\left[1+\frac{M_{j'}}{M_j}\right]  \frac{k_{2j}}{(1-e^2)^2} \left[\frac{R_j}{a}\right]^5,
\end{align}
and $j'$ refers to the other star.

\begin{figure}[ht!]
\epsscale{1.1}
\plotone{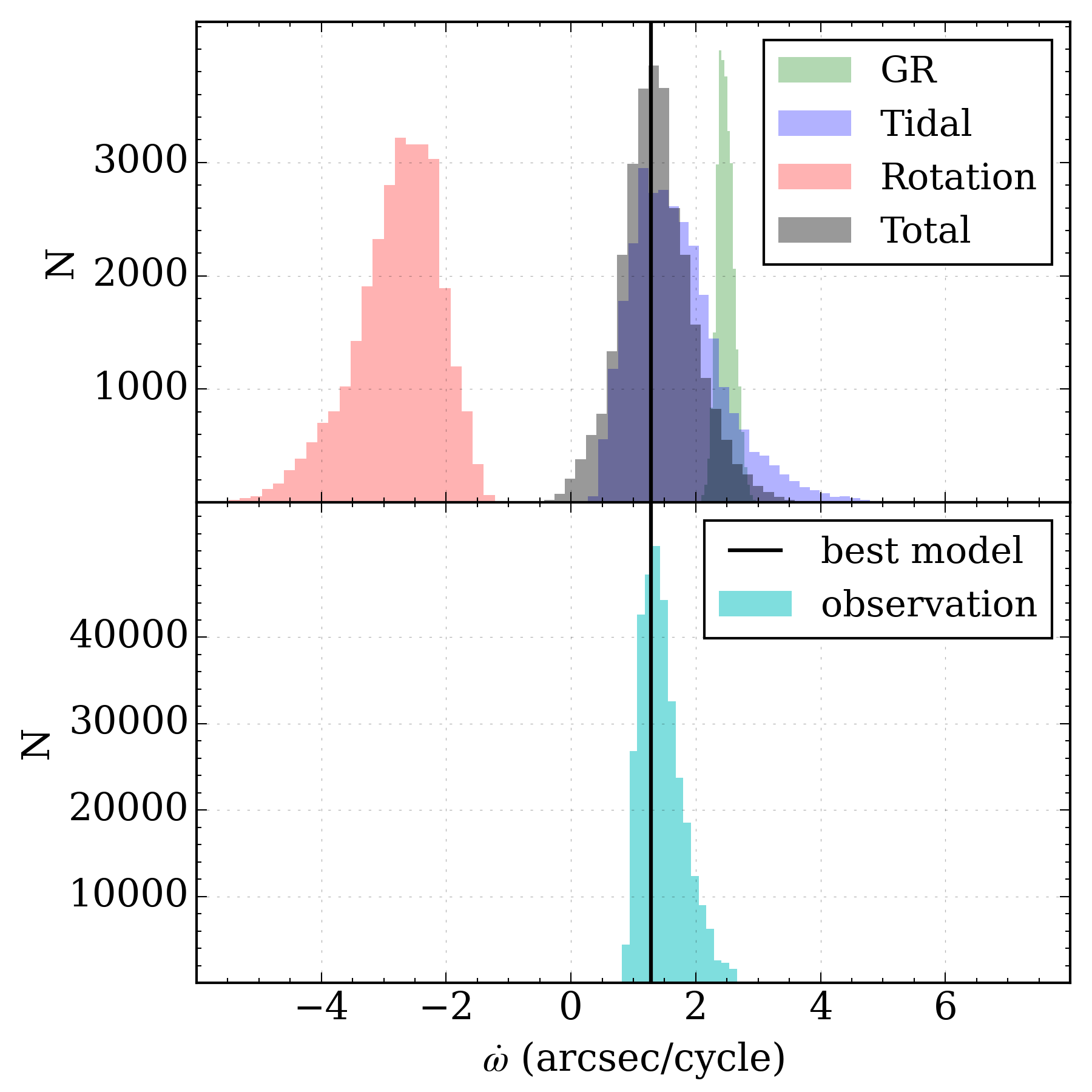}
\caption{ Top panel: Posterior probability distribution of for the theoretical
apsidal precession rate, based on the joint fit of the secondary eclipse light curve and previous spectroscopic measurements. The total predicted precession rate (gray) is composed of the general relativistic contribution (green), the tidal contribution (blue), and the rotational contribution (red). Bottom panel: Posterior probability distribution for the observed apsidal precession rate. The vertical line indicates the theoretical apsidal precession rate in the best-fit model.
\label{fig:wdot}}
\end{figure}

Given the posterior probability distributions for the orbital and spin parameters derived from our model of
the TESS light curve, and the spectroscopic measurements of $v\sin i$ and $\lambda$ (Table~\ref{tab:vsini}),
we evaluated Equation~\ref{eq:w1} to calculate the posteriors for the various contributions
to the theoretical apsidal precession rate.
The top panel of Figure~\ref{fig:wdot} shows these contributions,
along with the total.
The bottom panel of Figure~\ref{fig:wdot} shows the observed
apsidal precession rate, based on fitting the eclipse times.
The theoretical value of $1.35^{+0.58}_{-0.50}$ arcsec/cycle
agrees with the observed value of \hl{$1.41^{+0.39}_{-0.28}$} arcsec/cycle.
Compared with previous predictions \citep{claret2010di}, the theoretical
uncertainty has been reduced by approximately a factor of two,
mainly by excluding the long tail toward low values of $\dot\omega$
in the posterior probability density.


\begin{figure*}
\gridline{\fig{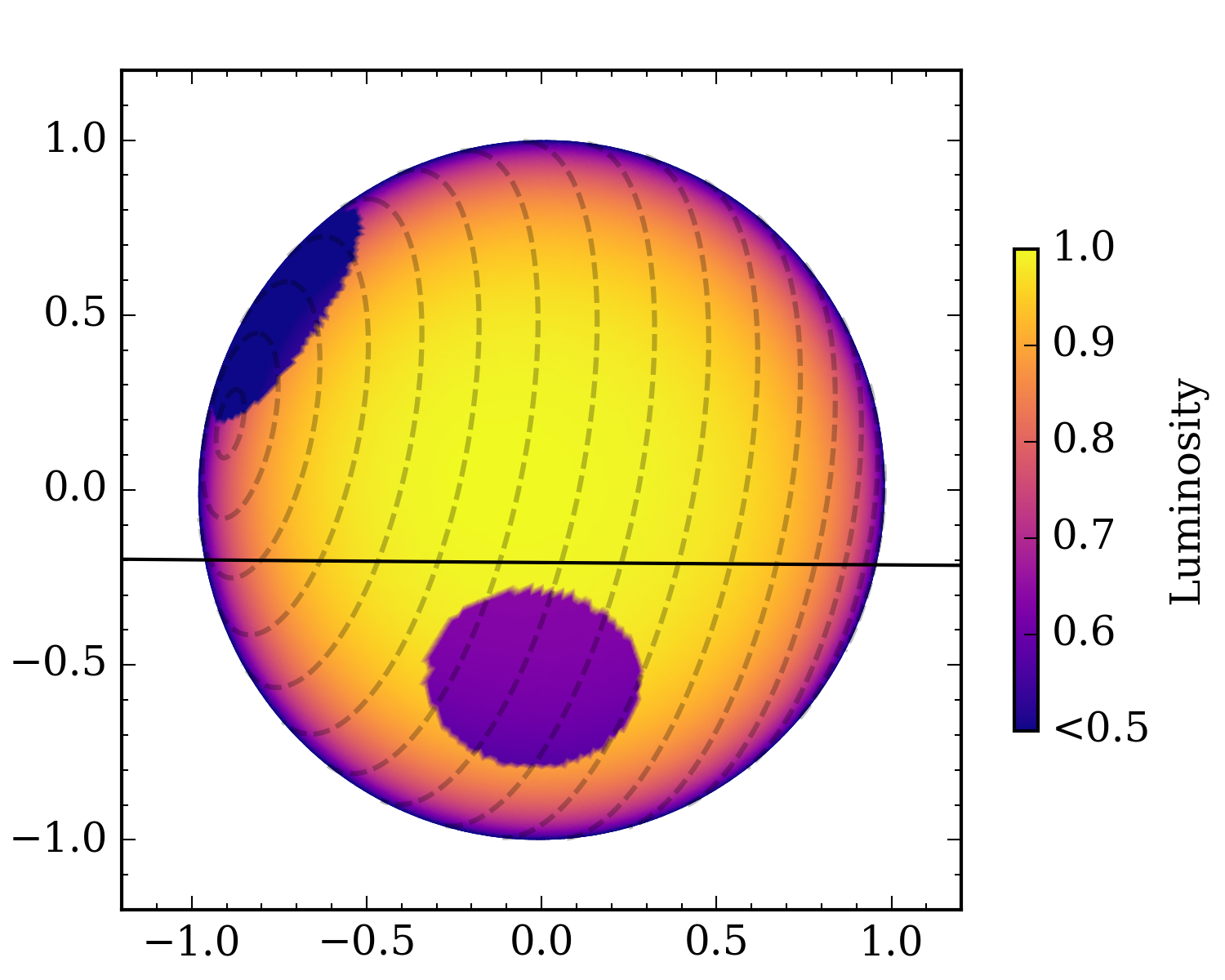}{0.5\textwidth}{(a) Primary star luminosity}
          \fig{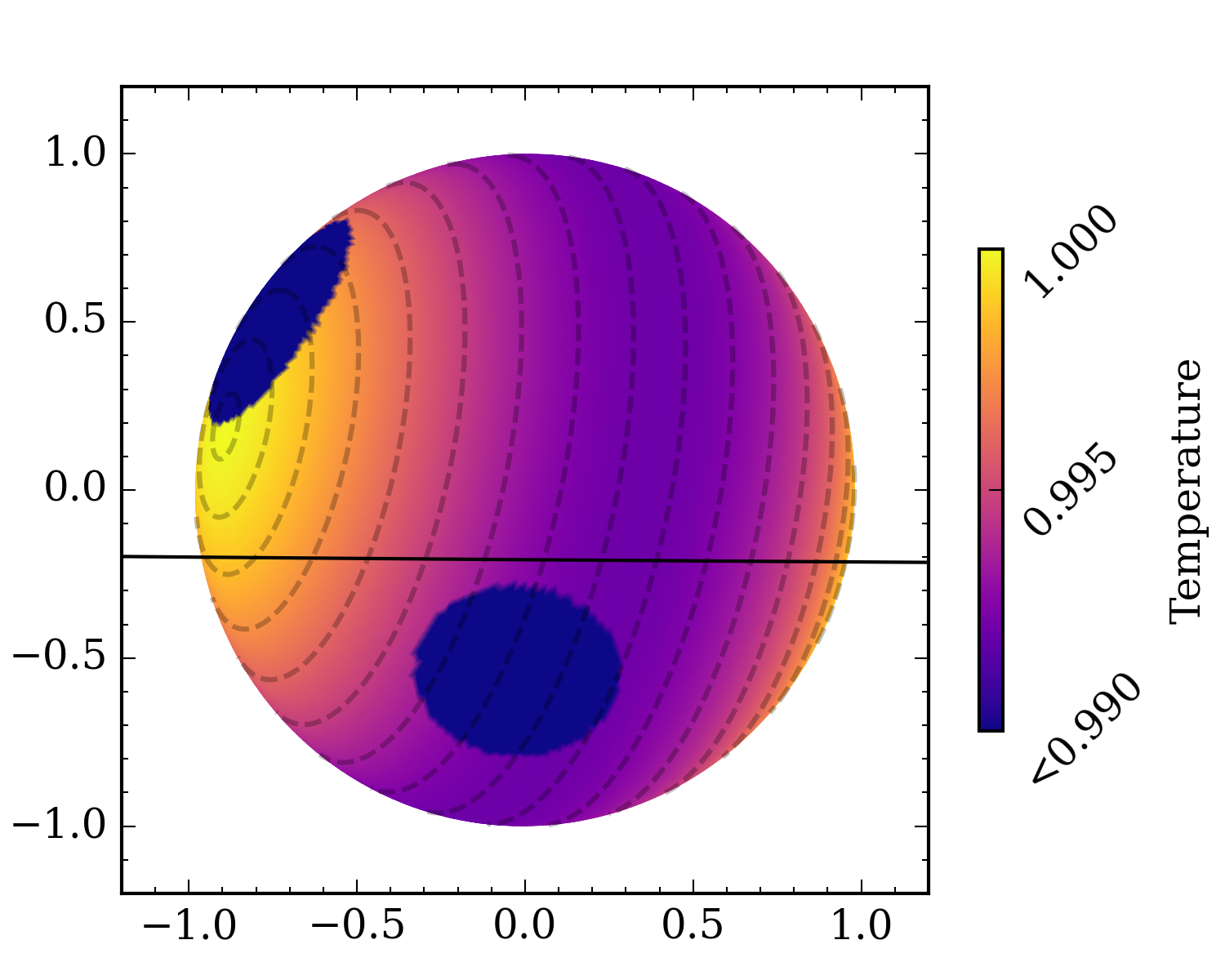}{0.5\textwidth}{(b) Primary star temperature}
          }
\gridline{\fig{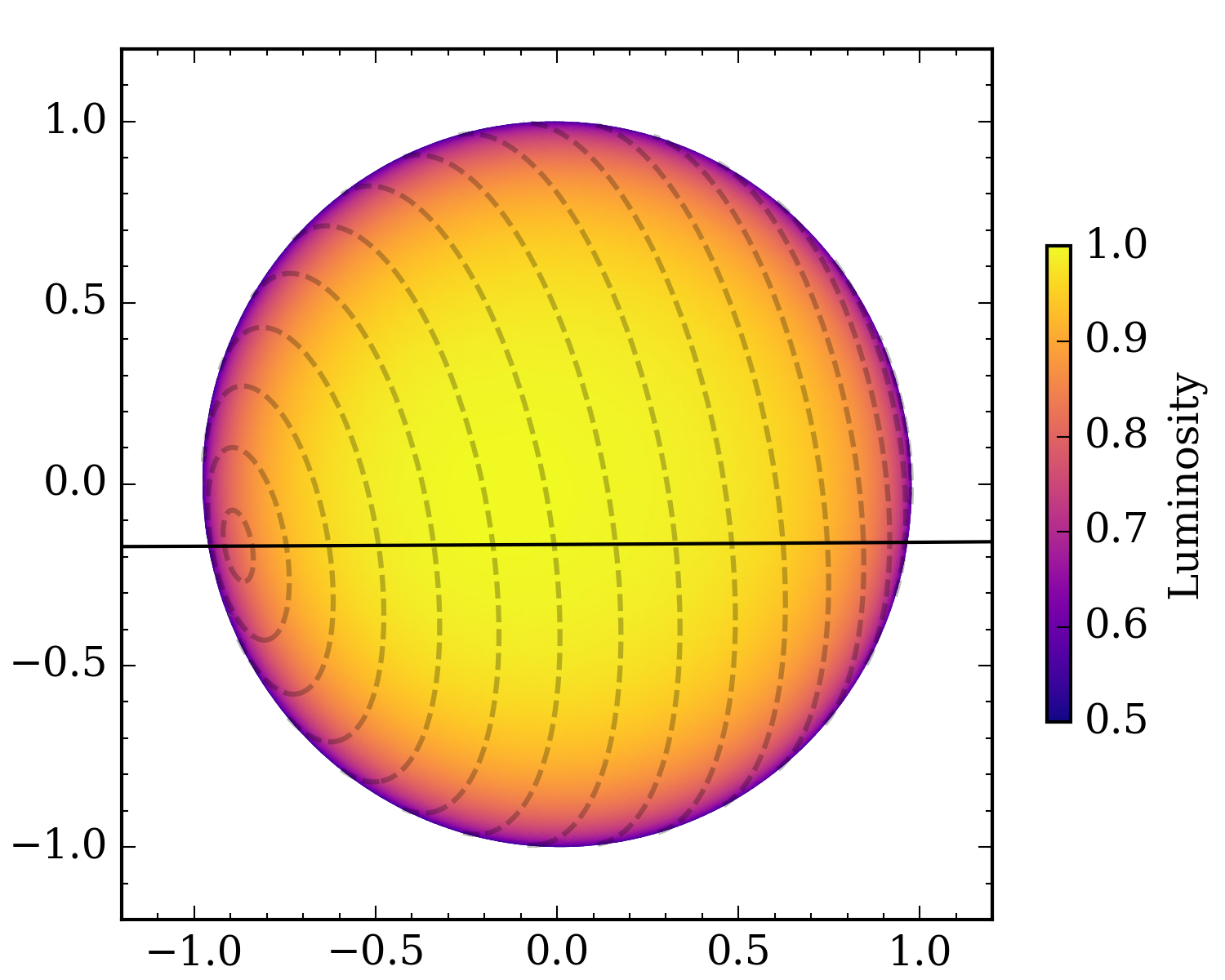}{0.5\textwidth}{(c) Secondary star luminosity}
          \fig{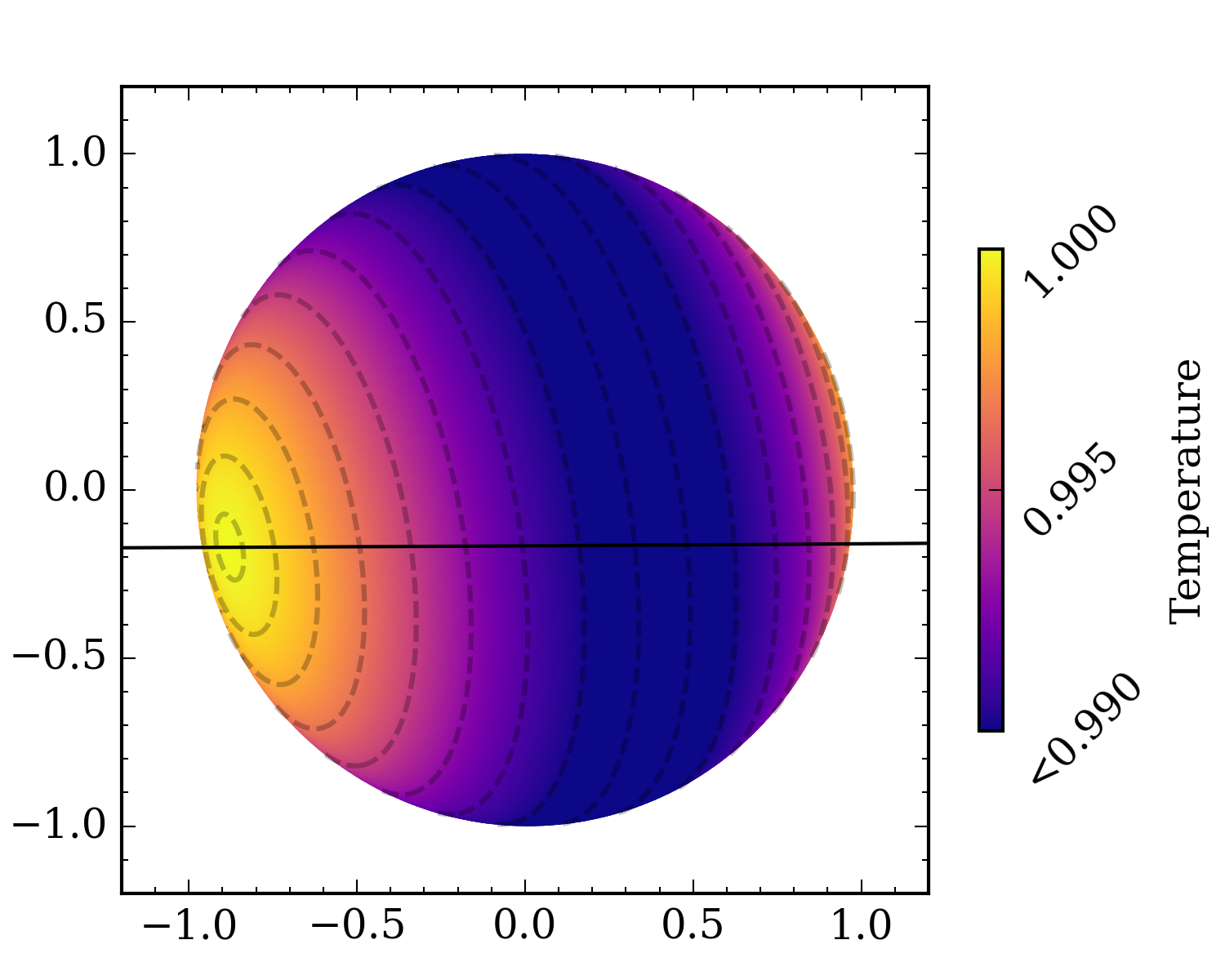}{0.5\textwidth}{(d) Secondary star temperature}
          }

\caption{ The luminosity distribution (a,c) and temperature distribution (b,d) for the primary and secondary stars, respectively. The black lines indicate the sky-plane trajectory of the foreground star during primary or secondary eclipses. Most of the variation in the emergent flux across the star is from limb darkening and spots. The much weaker variation with latitude is caused by gravity darkening, shown in panels (b) and (d).
\label{fig:LT}}
\end{figure*}

\section{Synthesis}
\label{sec:synthesis}

Given the good agreement between the observed and theoretical apsidal precession rates,
we proceeded to fit all the data together to arrive at the best possible determination
of the system parameters. We combined the $\chi^2$ contributions from the TESS secondary
eclipse light curve ($\chi^2_\text{lc}$), the spectroscopic $v\sin i$ and $\lambda$ measurements ($\chi^2_\text{spec}$), and the eclipse timings ($\chi^2_\text{timing}$). We minimized
\begin{align}
\label{eq:chi}
\chi^2_\text{tot}=\chi^2_\text{lc}+\chi^2_\text{spec}+\chi^2_\text{timing},
\end{align}
while also using Equation~\ref{eq:w1} to enforce agreement between
the observed and theoretical apsidal precession rates.

\begin{deluxetable}{lcr}
\tablenum{3}
\tablecaption{Orbital parameters of DI-Herculis \label{tab:param}}
\tablewidth{0pt}
\tablehead{
\colhead{Parameter} & \colhead{Posterior} & \colhead{Prior} }
\startdata
$R_1\ [R_\odot]$ & $2.80^{+0.10}_{-0.09}$ & \\
$R_2\ [R_\odot]$ & $2.58^{+0.09}_{-0.09}$ & \\
$M_1\ [M_\odot]$ & $5.1^{+0.2}_{-0.2}$ & $5.1\pm0.2$\\
$M_2\ [M_\odot]$ & $4.4^{+0.2}_{-0.2}$ & $4.4\pm0.2$\\
$i_o\ [^\circ]$ & $89.02^{+0.09}_{-0.11}$ & \\
$e$ & $0.51^{+0.02}_{-0.01}$ & \\
$\omega\ [^\circ]$ & $326^{+3}_{-3}$ & \\
$u_1$ & $0.13^{+0.04}_{-0.03}$ & $0.124\pm0.1$\\
$u_2$ & $0.35^{+0.07}_{-0.07}$ & $0.262\pm0.1$\\
$P_{rot,1}$ [day] & $1.10^{+0.06}_{-0.05}$ & $1.07\pm0.1$\\
$P_{rot,2}$ [day] & $0.97^{+0.05}_{-0.05}$ & \\
$i_{s,1}\ [^\circ]$ & $73^{+6}_{-6}$ & \\
$i_{s,2}\ [^\circ]$ & $109^{+8}_{-8}$ & \\
$\lambda_1\ [^\circ]$ & $-74^{+2}_{-3}$ & \\
$\lambda_2\ [^\circ]$ & $79^{+2}_{-3}$ & \\
$\psi_1\ [^\circ]$ & $75^{+3}_{-3}$ & \\
$\psi_2\ [^\circ]$ & $80^{+3}_{-3}$ & \\
$ \beta_{g,2}$ & $0.08^{+0.02}_{-0.02}$ & $0.086\pm0.025$\\
$T_{0,2}$ [BJD] & $2459016.24169^{+0.00003}_{-0.00003}$ & \\
$P_{orb}$ [day] & $10.55004^{+0.00002}_{-0.00002}$ & \\
 \hline
 \\[-1em]
  &Fixed parameters& \\
 \\[-1em]
 \hline
$T_{\text{eff},1}$ & $17300\pm800$ & \citep{claret2010di} \\
$T_{\text{eff},2}$ & $15400\pm800$ & \citep{claret2010di}\\
$\log k_{2,1}$ & $-2.146\pm0.050$ & \citep{Claret+2021}\\
$\log k_{2,2}$ & $-2.171\pm0.050$ &\citep{Claret+2021}\\
$\eta_1$ & $0.056$ &\citep{claret2019updating}\\
$\eta_2$ & $0.054$ &\citep{claret2019updating}\\
\hline
\enddata
\tablecomments{The orbital parameters are determined from the synthesized fit of TESS secondary eclipse lightcurve, $v\sin{i}$, $\lambda$, and archival eclipse timings. Due to spin and orbital precession, $i_o$, $\omega$, $i_{s,j}$, $\lambda_j$($j=1,2$) are functions of time. The tabulated values are valid for June 2020.}
\end{deluxetable}
The third column of Table~\ref{tab:param} summarizes the informative prior
constraints on the model.
The masses of both components were adopted from \cite{albrecht2009misaligned}.
The limb-darkening coefficients and the gravity-darkening coefficient
were taken from \cite{claret2017limb}. 
The rotation period of the primary star was determined in Section~\ref{sec:rot}.

The minimum-$\chi^2_\text{tot}$ model has \hl{$\chi^2_\text{lc}=556$} with 550 normalized flux data points, shown in Figure~\ref{fig:lc2}); \hl{$\chi^2_\text{spec}=2.9$} with 12 measurements of $v\sin i$ and $\lambda$ (Figure~\ref{fig:vsini}); and \hl{$\chi^2_\text{timing}=58$} with \hl{64} eclipse timings (Figure~\ref{fig:O-C}). 
The bottom two panels of 
Figure~\ref{fig:LT} show the
intensity and temperature
distribution across the secondary star in the best-fitting model.

We explored the 18-dimensional parameter space using the MCMC method
of \cite{foreman2013emcee}. We launched \hl{36} walkers near the maximum-likelihood model parameters, and sampled the posterior distribution through 88{,}000 iterations, resulting in $\sim\!3\times10^6$ samples. 
Table \ref{tab:param} gives the results, based on the marginalized posterior
probability distributions.
Although we did not place any informative priors on the eccentricity and argument
of pericenter, we note that our results are consistent with previous work by
\cite{popper1982rediscussion}, who found $e=0.489\pm0.003$ and $\omega=329.9\pm0.6$.

 
\section{Discussion}
\label{sec:concl}

We determined the three-dimensional spin orientation of the two stars of
DI Herculis, through a combination of (i) the TESS secondary eclipse light curve,
which showed gravity-darkening signatures, (ii) a measurement of the
rotation period of the primary star, and (iii) previous measurements of the
projected obliquities and projected rotation velocities.
This allowed us to reduce the fractional uncertainty in the theoretical
apsidal precession rate by about a factor of two.
We were also able to reduce the fractional uncertainty in the observed
rate by extending the time baseline over which mid-eclipse times
have been measured.
After confirming that the observed and theoretical rates are in agreement,
we used the theoretical apsidal precession rate as a prior constraint
in order to improve the determintaion of the other system parameters.

There appear to be at least two large spots on the primary star,
based on the out-of-eclipse stellar variability and spot-crossing anomalies
that were observed during eclipses. 
Although we are admittedly not experts in the variability of early-type
stars, we were surprised to see evidence for persistent dark spots
because we thought spot variability tends to be restricted to late-type stars with
magnetic activity cycles.
A literature search revealed that ``chemical spots'' have been inferred to exist on B-type stars
exhibiting peculiar HgMn abundances \citep{korhonen2013chemical},
but HgMn stars are characterized by slow rotational velocities ($v \sin{i}\leq 29$ km/s) \citep{hubrig2012magnetic},
and little or no photometric variability,
unlike the case of DI Herculis.
Magnetic Bp stars can exhibit photometric variability due to chemical spots,
but apparently such stars are very rarely found in close binaries \citep{kochukhov2018hd}. Among the known close binaries involving magnetic Bp stars, the system is either tidally locked \citep{kochukhov2018hd,shultz2015detection,shultz2019mobster} or has $v \sin{i}<20$ km/s \citep{landstreet2017bd}. It is suspected that the system must be synchronized for magnetic effects to dominate stellar variability \citep{pablo2019epsilon}. 

We did find a report of a HgMn binary with unexplained and possibly rotationally-induced $\sim$1 day photometric variability \citep{morel2014search}, and a magnetic Bp binary with weakly detected $\sim$0.9 day light curve modulation \citep{bolton1998hd}. In addition, \cite{shultz2019nu} reported a rapidly rotating magnetic B-type star in hierarchical triple system,
with $v \sin{i}\approx200$ km/ and $P_{rot}=1.09$~days. Perhaps these exceptional cases are
somehow related to the spots of DI Herculis.

The reason for the high obliquities is still unknown.  Possible explanations include primordial misalignment \citep{philippov2013analysis}, the disappearance of an inclined circumbinary disk \citep{anderson2020excitation}, and obliquity excitation by a distant third star \citep{albrecht2009misaligned,philippov2013analysis,claret2010di}. It is recognized that the Lidov–Kozai mechanism can drive the obliquities of the inner binary when the orbits of the inner binary and the third companion are mutually inclined \citep{lidov1962evolution,kozai1962secular}. \cite{anderson2017eccentricity} investigated the requirements for obliquity excitation in stellar triples, and suggested that a third companion needed to be sufficiently massive and close (within several AU) to excite the obliquities of DI Herculis. Subsequently, Anderson \& Winn (submitted) found that such a nearby massive companion can essentially
be ruled out with current radial-velocity and imaging data. 

On a practical note, we found that calculating the irreducible antisymmetric component
of the eclipse light-curve was a useful and simple diagnostic for the effects of
gravity darkening and misalignment on the TESS light curve.
It can be computed rapidly, without detailed parametric modeling.
We intend to search for antisymmetry in all of the high-signal-to-noise
TESS eclipse light curves for well-detached and non-synchronized binaries,
as a means of identifying other
misaligned binaries.  The search may also reveal other physical effects that produce
antisymmetric signals, such as high eccentricities and the photometric Rossiter-McLaughlin
effect.

\begin{acknowledgments}
The HARPS-N data reported in this paper were obtained through program
A33TAC\_11 and are archived in the INAF (Istituto Nazionale di Astrofisica) Science Archive (https://www.ia2.inaf.it/).
Funding for the Stellar Astrophysics Centre is provided by The Danish National Research Foundation (Grant agreement no.: DNRF106).
\end{acknowledgments}

%

\vspace{5mm}
\facilities{TESS,TNG}


\software{PyTransit \citep{Parviainen2015},
          emcee \citep{foreman2013emcee}
          }





\bibliography{main}{}

\begin{thebibliography}{}
\expandafter\ifx\csname natexlab\endcsname\relax\def\natexlab#1{#1}\fi
\providecommand{\url}[1]{\href{#1}{#1}}
\providecommand{\dodoi}[1]{doi:~\href{http://doi.org/#1}{\nolinkurl{#1}}}
\providecommand{\doeprint}[1]{\href{http://ascl.net/#1}{\nolinkurl{http://ascl.net/#1}}}
\providecommand{\doarXiv}[1]{\href{https://arxiv.org/abs/#1}{\nolinkurl{https://arxiv.org/abs/#1}}}

\bibitem[{Aerts {et~al.}(2006)Aerts, De~Cat, Kuschnig, Matthews, Guenther,
  Moffat, Rucinski, Sasselov, Walker, \& Weiss}]{aerts2006discovery}
Aerts, C., De~Cat, P., Kuschnig, R., {et~al.} 2006, The Astrophysical Journal
  Letters, 642, L165

\bibitem[{Albrecht {et~al.}(2009)Albrecht, Reffert, Snellen, \&
  Winn}]{albrecht2009misaligned}
Albrecht, S., Reffert, S., Snellen, I.~A., \& Winn, J.~N. 2009, Nature, 461,
  373

\bibitem[{Anderson \& Lai(2020)}]{anderson2020excitation}
Anderson, K.~R., \& Lai, D. 2020, The Astrophysical Journal, 906, 17

\bibitem[{Anderson {et~al.}(2017)Anderson, Lai, \&
  Storch}]{anderson2017eccentricity}
Anderson, K.~R., Lai, D., \& Storch, N.~I. 2017, Monthly Notices of the Royal
  Astronomical Society, 467, 3066

\bibitem[{Barker \& O'Connell(1975)}]{barker1975gravitational}
Barker, B.~M., \& O'Connell, R.~F. 1975, Physical Review D, 12, 329

\bibitem[{{Barnes}(2009)}]{Barnes2009}
{Barnes}, J.~W. 2009, \apj, 705, 683, \dodoi{10.1088/0004-637X/705/1/683}

\bibitem[{Bolton {et~al.}(1998)Bolton, Harmanec, Lyons, Odell, \&
  Pyper}]{bolton1998hd}
Bolton, C., Harmanec, P., Lyons, R., Odell, A., \& Pyper, D.~M. 1998, Astronomy
  and Astrophysics, 337, 183

\bibitem[{Claret(1998)}]{claret1998some}
Claret, A. 1998, Astronomy and Astrophysics, 330, 533

\bibitem[{Claret(2017)}]{claret2017limb}
---. 2017, Astronomy \& Astrophysics, 600, A30

\bibitem[{Claret(2019)}]{claret2019updating}
---. 2019, Astronomy \& Astrophysics, 628, A29

\bibitem[{{Claret} {et~al.}(2021){Claret}, {Gim{\'e}nez}, {Baroch}, {Ribas},
  {Morales}, \& {Anglada-Escud{\'e}}}]{Claret+2021}
{Claret}, A., {Gim{\'e}nez}, A., {Baroch}, D., {et~al.} 2021, arXiv e-prints,
  arXiv:2107.10765.
\newblock \doarXiv{2107.10765}

\bibitem[{Claret {et~al.}(2010)Claret, Torres, \& Wolf}]{claret2010di}
Claret, A., Torres, G., \& Wolf, M. 2010, Astronomy \& Astrophysics, 515, A4

\bibitem[{Conroy {et~al.}(2020)Conroy, Kochoska, Hey, Pablo, Hambleton, Jones,
  Giammarco, Abdul-Masih, \& Pr{\v{s}}a}]{conroy2020physics}
Conroy, K.~E., Kochoska, A., Hey, D., {et~al.} 2020, The Astrophysical Journal
  Supplement Series, 250, 34

\bibitem[{Foreman-Mackey {et~al.}(2013)Foreman-Mackey, Hogg, Lang, \&
  Goodman}]{foreman2013emcee}
Foreman-Mackey, D., Hogg, D.~W., Lang, D., \& Goodman, J. 2013, Publications of
  the Astronomical Society of the Pacific, 125, 306

\bibitem[{Gimenez \& Garcia-Pelayo(1983)}]{gimenez1983new}
Gimenez, A., \& Garcia-Pelayo, J.~M. 1983, Astrophysics and Space Science, 92,
  203

\bibitem[{Guinan \& Maloney(1985)}]{guinan1985apsidal}
Guinan, E.~F., \& Maloney, F.~P. 1985, The Astronomical Journal, 90, 1519

\bibitem[{Hubrig {et~al.}(2012)Hubrig, Gonzalez, Ilyin, Korhonen, Sch{\"o}ller,
  Savanov, Arlt, Castelli, Curto, Briquet, {et~al.}}]{hubrig2012magnetic}
Hubrig, S., Gonzalez, J.~F., Ilyin, I., {et~al.} 2012, Astronomy \&
  Astrophysics, 547, A90

\bibitem[{Kochukhov {et~al.}(2018)Kochukhov, Johnston, Alecian, Wade, \&
  Collaboration}]{kochukhov2018hd}
Kochukhov, O., Johnston, C., Alecian, E., Wade, G., \& Collaboration, B. 2018,
  Monthly Notices of the Royal Astronomical Society, 478, 1749

\bibitem[{Korhonen {et~al.}(2013)Korhonen, Gonzalez, Briquet, Soriano, Hubrig,
  Savanov, Hackman, Ilyin, Eulaers, \& Pessemier}]{korhonen2013chemical}
Korhonen, H., Gonzalez, J.~F., Briquet, M., {et~al.} 2013, Astronomy \&
  Astrophysics, 553, A27

\bibitem[{Kozai(1962)}]{kozai1962secular}
Kozai, Y. 1962, The Astronomical Journal, 67, 591

\bibitem[{Kozyreva \& Bagaev(2009)}]{kozyreva2009pulsations}
Kozyreva, V.~S., \& Bagaev, L.~A. 2009, Astronomy letters, 35, 483

\bibitem[{Landstreet {et~al.}(2017)Landstreet, Kochukhov, Alecian, Bailey,
  Mathis, Neiner, \& Wade}]{landstreet2017bd}
Landstreet, J., Kochukhov, O., Alecian, E., {et~al.} 2017, Astronomy \&
  Astrophysics, 601, A129

\bibitem[{Lidov(1962)}]{lidov1962evolution}
Lidov, M. 1962, Planetary and Space Science, 9, 719

\bibitem[{Martynov \& Khaliullin(1980)}]{martynov1980relativistic}
Martynov, D.~Y., \& Khaliullin, K.~F. 1980, Astrophysics and Space Science, 71,
  147

\bibitem[{Masuda(2015)}]{masuda2015spin}
Masuda, K. 2015, The Astrophysical Journal, 805, 28

\bibitem[{{McQuillan} {et~al.}(2013){McQuillan}, {Aigrain}, \&
  {Mazeh}}]{2013MNRAS.432.1203M}
{McQuillan}, A., {Aigrain}, S., \& {Mazeh}, T. 2013, \mnras, 432, 1203,
  \dodoi{10.1093/mnras/stt536}

\bibitem[{Morel {et~al.}(2014)Morel, Briquet, Auvergne, Alecian, Ghazaryan,
  Niemczura, Fossati, Lehmann, Hubrig, Ulusoy, {et~al.}}]{morel2014search}
Morel, T., Briquet, M., Auvergne, M., {et~al.} 2014, Astronomy \& Astrophysics,
  561, A35

\bibitem[{Pablo {et~al.}(2019)Pablo, Shultz, Fuller, Wade, Paunzen, Mathis,
  Le~Bouquin, Pigulski, Handler, Alecian, {et~al.}}]{pablo2019epsilon}
Pablo, H., Shultz, M., Fuller, J., {et~al.} 2019, Monthly Notices of the Royal
  Astronomical Society, 488, 64

\bibitem[{Parviainen(2015)}]{Parviainen2015}
Parviainen, H. 2015, MNRAS, 450, 3233, \dodoi{10.1093/mnras/stv894}

\bibitem[{Philippov \& Rafikov(2013)}]{philippov2013analysis}
Philippov, A.~A., \& Rafikov, R.~R. 2013, The Astrophysical Journal, 768, 112

\bibitem[{Popper(1982)}]{popper1982rediscussion}
Popper, D. 1982, The Astrophysical Journal, 254, 203

\bibitem[{{Reisenberger} \& {Guinan}(1989)}]{1989AJ.....97..216R}
{Reisenberger}, M.~P., \& {Guinan}, E.~F. 1989, \aj, 97, 216,
  \dodoi{10.1086/114972}

\bibitem[{{Ricker} {et~al.}(2015){Ricker}, {Winn}, {Vanderspek}, {Latham},
  {Bakos}, {Bean}, {Berta-Thompson}, {Brown}, {Buchhave}, {Butler}, {Butler},
  {Chaplin}, {Charbonneau}, {Christensen-Dalsgaard}, {Clampin}, {Deming},
  {Doty}, {De Lee}, {Dressing}, {Dunham}, {Endl}, {Fressin}, {Ge}, {Henning},
  {Holman}, {Howard}, {Ida}, {Jenkins}, {Jernigan}, {Johnson}, {Kaltenegger},
  {Kawai}, {Kjeldsen}, {Laughlin}, {Levine}, {Lin}, {Lissauer}, {MacQueen},
  {Marcy}, {McCullough}, {Morton}, {Narita}, {Paegert}, {Palle}, {Pepe},
  {Pepper}, {Quirrenbach}, {Rinehart}, {Sasselov}, {Sato}, {Seager},
  {Sozzetti}, {Stassun}, {Sullivan}, {Szentgyorgyi}, {Torres}, {Udry}, \&
  {Villasenor}}]{Ricker+2015}
{Ricker}, G.~R., {Winn}, J.~N., {Vanderspek}, R., {et~al.} 2015, Journal of
  Astronomical Telescopes, Instruments, and Systems, 1, 014003,
  \dodoi{10.1117/1.JATIS.1.1.014003}

\bibitem[{{Shakura}(1985)}]{Shakura1985}
{Shakura}, N.~I. 1985, Soviet Astronomy Letters, 11, 224

\bibitem[{Shultz {et~al.}(2015)Shultz, Wade, Alecian, \&
  Collaboration}]{shultz2015detection}
Shultz, M., Wade, G., Alecian, E., \& Collaboration, B. 2015, Monthly Notices
  of the Royal Astronomical Society: Letters, 454, L1

\bibitem[{Shultz {et~al.}(2019{\natexlab{a}})Shultz, Johnston, Labadie-Bartz,
  Petit, David-Uraz, Kochukhov, Wade, Pepper, Stassun, Rodriguez,
  {et~al.}}]{shultz2019mobster}
Shultz, M., Johnston, C., Labadie-Bartz, J., {et~al.} 2019{\natexlab{a}},
  Monthly Notices of the Royal Astronomical Society, 490, 4154

\bibitem[{Shultz {et~al.}(2019{\natexlab{b}})Shultz, Le~Bouquin, Rivinius,
  Wade, Kochukhov, Alecian, Petit, Pfuhl, Karl, Gao, {et~al.}}]{shultz2019nu}
Shultz, M., Le~Bouquin, J., Rivinius, T., {et~al.} 2019{\natexlab{b}}, Monthly
  Notices of the Royal Astronomical Society, 482, 3950

\bibitem[{Von~Zeipel(1924)}]{von1924radiative}
Von~Zeipel, H. 1924, Monthly Notices of the Royal Astronomical Society, 84, 665

\end{thebibliography}
\bibliographystyle{aasjournal}



\end{document}